\documentclass[english,11pt,a4paper]{article}
\pdfoutput=1
\usepackage{jheppub}

\usepackage[T1]{fontenc}
\usepackage[font=small,labelfont=bf]{caption}
\usepackage{enumitem}
\usepackage[force]{feynmp-auto}
\usepackage{hyperref}
\usepackage{cancel}
\usepackage{multirow}
\usepackage{booktabs}
\usepackage{subcaption}
\usepackage{url}
\usepackage{tikz}
\usepackage{adjustbox}
\usepackage{array}
\usepackage{makecell}
\usepackage{breqn}
\usepackage{tikz-feynman}
\usetikzlibrary{positioning}

\newcommand{\eg}{{\em e.g.}}
\newcommand{\ie}{{\em i.e.}}

\renewcommand{\eqref}[1]{Eq.~(\ref{eqn:#1})}
\newcommand{\eqsref}[2]{Eqs.~(\ref{eqn:#1}) and (\ref{eqn:#2})}

\newcommand{\secref}[1]{Sec.~\ref{sec:#1}}

\newcommand{\appref}[1]{Appendix~\ref{sec:#1}}
\newcommand{\figref}[1]{Fig.~\ref{fig:#1}}

\newcommand{\tableref}[1]{Table~\ref{tab:#1}}

\newcommand{\beqa}{\begin{eqnarray}}
\newcommand{\eeqa}{\end{eqnarray}}
\newcommand{\beq}{\begin{equation}}
\newcommand{\eeq}{\end{equation}}
\newcommand{\nn}{\nonumber}

\newcommand{\tr}{{\rm Tr}}
\newcommand{\cm}{{\cal M}}

\newcommand{\bpm}{\begin{pmatrix}}
\newcommand{\epm}{\end{pmatrix}}
\newcommand{\ket}[1]{{\left|#1\right\rangle}}
\newcommand{\bs}[1]{{\boldsymbol{#1}}}
\newcommand{\bra}[1]{{\left\langle #1\right|}}
\newcommand{\Ket}[1]{{\left|#1 \right] }}
\newcommand{\Bra}[1]{{\left[ #1\right|}}

\newcommand{\parn}[1]{{\left(#1\right) }}
\newcommand{\parnS}[1]{{\left[#1 \right]  }}

\newcommand{\sqb}[1]{\left[ #1 \right] }
\newcommand{\anb}[1]{\left\langle #1 \right\rangle  }

\newcommand{\braKet}[3]{\left\langle #1|#2|#3\right]}

\newcommand{\eqCS}{\,,~~}

\newcommand{\splitFunc}[1]{\mathrm{Split}}

\newcommand{\Tr}{\mathop{\mathrm{Tr}}}

\newcommand{\SUN}[1]{\mathrm{SU}\left(#1\right)}
\newcommand{\SUNlr}[2]{\mathrm{SU}_{#2}\left(#1\right)}

\newcommand{\SL}[2]{\mathrm{SL}\left(#1,#2\right)}
\newcommand{\SD}[1]{#1_{\mathrm{SD}}}
\newcommand{\ASD}[1]{#1_{\mathrm{ASD}}}
\newcommand{\GSD}{G_{\mathrm{SD}}}
\newcommand{\GASD}{G_{\mathrm{ASD}}}

\makeatletter
\def\EE{\@ifnextchar-{\@@EE}{\@EE}}
\def\@EE#1{\ifnum#1=1 \times10 \else \times10^{#1}\fi}
\def\@@EE#1#2{\times10^{-#2}}
\makeatother

\title{Effective Field Theory Amplitudes the On-Shell Way:
  Scalar and Vector Couplings to  Gluons}

\author{Yael Shadmi}
\author{and Yaniv Weiss}
\affiliation{Physics Department, Technion---Israel Institute of Technology,\\ Haifa 3200003, Israel}

\emailAdd{yshadmi@physics.technion.ac.il}
\emailAdd{yanivwe@campus.technion.ac.il}

\abstract{%
  We use on-shell methods to calculate tree-level effective field theory
  (EFT) amplitudes, with no reference to the EFT operators.
  Lorentz symmetry, unitarity and Bose statistics determine the allowed 
  kinematical structures.
    As a by-product, the number of independent EFT operators simply follows from
  the set of polynomials in the Mandelstam invariants, subject to
  kinematical constraints.
  We demonstrate this approach by calculating several amplitudes with
  a massive, SM-singlet, scalar ($h$) or vector ($Z^\prime$) particle
  coupled to gluons. Specifically, we calculate $hggg$, $hhgg$
  and $Z^\prime ggg$ amplitudes, which are
  relevant for the LHC production and three-gluon decays of
  the massive particle.
  We then use the results to derive the massless-$Z^\prime$ amplitudes, and show how the massive amplitudes
  decompose into the massless-vector plus scalar amplitudes.
Amplitudes with the gluons replaced by photons are  straightforwardly obtained from the above. 
}

\begin{document}
\maketitle


\section{Introduction and outline} \label{sec:intro}
Effective field theories (EFTs) provide a systematic and largely
model-independent parameterization of beyond the standard model (BSM) effects.
They have taken center-stage recently, in response to the 
lack of hints for BSM physics at LHC experiments,
and the attendant, and welcome, theoretical
uncertainty as to the form BSM physics might take.
EFTs are well motivated and in principle, straightforward to use.
In practice however, their application is quite involved.
Non-renormalizable operators give rise to vertices with large numbers of external
legs and/or derivatives, adding to the complexity of Feynman-diagrammatic
gauge theory calculations.
Furthermore, the very first step in an EFT calculation is the identification of
the full set of independent operators.
This has been the subject of intense study in recent
years~\cite{Jenkins:2009dy,Grzadkowski:2010es,Lehman:2015coa,Henning:2015alf,Lehman:2015via,Henning:2017fpj, Gripaios:2018zrz}.

Ultimately, however,  the quantities of interest are
physical observables, obtained from  on-shell amplitudes.
When calculating these observables, the inter-dependencies of various
EFT operators should be manifest.
Thus,  we propose to bypass EFT Lagrangians altogether, and quantify possible
deviations from the SM directly at the level of physical scattering amplitudes.
We consider a low-energy theory below some scale $\Lambda$.
The amplitudes in this low-energy theory are constrained by Lorentz symmetry,
unitarity, and Bose or Fermi statistics. These determine the different kinematical 
structures that can appear at any dimension, accompanied by the appropriate negative
power of the scale $\Lambda$.
Each of these structures appears with some unknown numerical coefficient.
From the Lagrangian point of view,
the number of these unknown parameters equals the number of independent
operators at a given dimension. Thus, a by-product of  
the on-shell computation
is a counting of independent EFT operators.
This counting is extremely simple, since we are counting  independent terms
in polynomials of the kinematic invariants, $s_{ij}= (p_i+p_j)^2$, subject to
kinematical constraints, as opposed to the Hilbert series associated with
polynomials of operators (see also~\cite{Cheung:2016drk,Henning:2017fpj}).

We apply on-shell methods to calculate EFT amplitudes in two scenarios.
In the first, we augment the SM by a massive spin-0, gauge singlet, $h$,
with couplings to gluons,  and consider an EFT comprised of $h$ and
the gluons\footnote{$h$ can be thought of either as the SM Higgs,
  or as a new spin-0 particle.}.
We calculate tree-level amplitudes with one scalar  and three gluons,
as well as amplitudes with two scalars  and two gluons.
Thus we reproduce some of the results of~\cite{DelDuca:2004wt,Dixon:2004za},
where Higgs plus $n$-gluon amplitudes were derived for the SM dimension-5
top loop operator. However, we are interested in the most general EFT coupling
the scalar to the gluons, and our derivation captures the contributions of
additional higher-dimension operators, up to dimension 13.
For our second example, we consider a spin-1, gauge singlet,
$Z^\prime$, which couples to gluons,
and calculate the vector plus three gluon amplitude, $\cm(Z^\prime; ggg)$.
We then take the massless limit to derive the amplitudes with a massless $Z^\prime$.
Throughout, we ignore quarks.

Let us sketch our approach using the single $h$ plus three gluon amplitude,
$\cm(h;ggg)$, as an example.
Little group considerations determine the dependence of the amplitude on
spinor-products associated with the gluon momenta $p_{i=1,2,3}$.
The spinor-product factor can be multiplied by some
analytic function of the $s_{ij}$'s and $\Lambda$, $f(s_{ij};\Lambda)$.
At tree level, the only possible
structures that can appear in $f(s_{ij};\Lambda)$ are single-particle poles
in the $s_{ij}$'s plus non-negative powers
of the $s_{ij}$.
The former constitute the factorizable part of the amplitude, and are completely determined by the relevant three-point amplitudes.
In the case at hand, these are the two possible three-gluon helicity amplitudes,
and $\cm(h;gg)$.
Thus, they introduce three independent
parameters: one for $\cm(hgg)$, where only equal gluon helicities can appear
(with the convention that all external particles are incoming),
one for the $++-$ three-gluon amplitude, corresponding to the usual QCD coupling,
and one for the $+++$ three-gluon
amplitude,  generated by the dimension-six gluonic operator $\tr (G^3)$.
The non-negative powers of the $s_{ij}$'s arise from
non-renormalizable operators involving a single $h$ and three gluons.
These constitute the non-factorizable part of the amplitude, which is our focus here.
Symmetrizing over same-helicity gluons and  imposing the
kinematical constraints eliminates many of the possible structures.
The remaining terms appear with unknown coefficients. These correspond to the Wilson coefficients
of EFT operators, and their number gives the number of independent operators.

Alternatively, one could derive the amplitude by writing the most general ansatz
for the function $f(s_{ij};\Lambda)$, and requiring the correct factorization
in the different possible collinear limits~\cite{Dixon:2013uaa}.
With complex momenta, this correctly captures all the relevant three-point amplitudes, including those that vanish
on-shell.
In particular, it is easy to see that 
the only collinear singularities arise from the
QCD three-gluon couplings,
while higher-dimension operators do not introduce
any collinear singularities.
This generalizes the dimensional-analysis argument of~\cite{Dixon:1993xd}
that the operator $\tr(G^3)$ does not introduce any collinear singularities.

As mentioned above, the $++-$ three-gluon amplitude is associated with the
renormalizable QCD coupling, while the $+++$ amplitude corresponds to a $1/\Lambda^2$ coupling.
This behavior is generic: at leading order, amplitudes of different net helicities
are generated
 by operators of different dimensionality~\cite{Cheung:2015aba,Azatov:2016sqh}.
We will see that this fact emerges very simply from little group
and dimensional analysis considerations.
Thus, by constructing the full set of helicity amplitudes associated with a
given process, one essentially scans over the full set of operators of interest.

To illustrate the counting of operators in this approach, consider an
even simpler example, with a massless real scalar $\phi$, with a quartic
coupling but no cubic coupling. We can obtain the number of independent higher-dimension operators containing four $\phi$'s by considering the 4-$\phi$ amplitude,
\begin{align}
    &{\cal M}(p_1,p_2,p_3,p_4) = c_4 + c_6\,
	\frac{\left(s_{12}+\cdots\right)}{\Lambda^2} +
	c_8\,\frac{\left(s_{12}^2+\cdots\right)}{\Lambda^4} +
	c_8^\prime\, \frac{\left(s_{12} s_{13}+\cdots\right)}{\Lambda^4}
	+\cdots\,,
\end{align}
where the $c$'s are dimensionless coefficients, and the ellipses stand
for symmetric permutations. Because of momentum conservation, the term with $c_6$ actually vanishes, and we can trade the last term for the term with $c_8$.
Thus we see that there is one operator
at dimension-8, and no operator at dimension-6. 
Indeed, at dimension-6, $(\partial\phi)^2 \phi^2$ can be eliminated by a field redefinition. 
At dimension-8, the only independent operator is $(\partial\phi)^4$.

As a check of our results, we used the Mathematica
notebook of~\cite{Henning:2015alf} to derive the EFT Lagrangian for the scalar
plus gluon case, as well as for the massless-vector plus gluon case.
Our counting of operators
indeed reproduces the results of~\cite{Henning:2015alf}.


We stress that we do not introduce any new tools, but rather make use
of well-known on-shell methods (for reviews of the subject,
see~\eg~\cite{Dixon:2013uaa,Elvang:Book15,Cheung:2017pzi,Schwartz:Book2014}).
Gauge symmetry is an output, rather than an input, in
the on-shell approach.
This is evident in the counting of independent parameters in our examples,
which matches the number of independent gauge-invariant
operators. The amplitudes ``know about'' gauge invariance to all orders
by virtue of the aforementioned principles: Lorentz invariance, unitarity and
Bose statistics.

This paper is organized as follows. 
In~\secref{spin0} we calculate the $hgg$, $hggg$ and $hhgg$ amplitudes, and discuss the correspondence with the EFT operators.
In~\secref{spin1} we calculate the $Z^\prime ggg$ amplitudes, starting
from a massive $Z^\prime$, and then taking the high energy limit to obtain the
massless case. The massive-vector amplitudes split into different massless-$Z^\prime$ helicity amplitudes
plus scalar amplitudes. We discuss the correspondence with EFT operators, and the number of independent operators, in both the
massive and massless case.
\appref{notation} summarizes the essentials of the spinor formalism we use, for both  massless and massive particles.
Some details of the high-energy limit appear in~\appref{helimit}.

\section{Scalar plus gluon amplitudes}\label{sec:spin0}
Here we consider amplitudes with a single spin-0, gauge singlet, $h$,
and two or three gluons. We neglect quarks throughout this discussion.
Tree-level amplitudes with a single Higgs and any number of gluons were derived
in~\cite{DelDuca:2004wt,Dixon:2004za}, assuming the SM dimension-5 top-loop operator
\beq\label{eqn:smhiggs}
\frac1\Lambda h\, G^{\mu\nu} G_{\mu\nu}\,.
\eeq
Some of our analysis reproduces these known results.
Our aim however, is to generalize these results beyond the operator~\eqref{smhiggs}, to any possible higher-dimension operator, suppressed by the appropriate power of a single scale $\Lambda$.
The contribution of dimension-7 operators was inferred from Lorentz symmetry considerations in~\cite{Dawson:2014ora,Dawson:2015gka}.

\subsection{The scalar plus  two gluon amplitude~${\cal M}(h;gg)$}

We start with the single scalar, 2-gluon amplitude,
$\mathcal{M}\parn{h; g^{a, \, h_{1}}(p_1)g^{b , \,h_{2}}(p_2)}$.
The most general ansatz for this amplitude is,
\begin{equation}\label{eqn:higgs_2g_ansatz}
  \mathcal{M}\parn{h; g^{a, \, h_{1}}(p_1)\,g^{b , \,h_{2}}(p_2)} = \delta^{ab} \, \left[12\right]^n \,  f_{-\ell}\left(s_{12};\Lambda\right)\,,
\end{equation}
where $\delta^{ab}$ is a color factor, $h_1$, $h_2$ are the gluon helicities, $n$ is an integer,
$f_{-\ell}$ is an analytic function of mass dimension $-\ell$,
and $s_{12}=\left(p_{1}+p_{2}\right)^{2}=m^2$.
Since $h$ is a scalar, the only little group weights are carried by the gluon
spinor products.
We then have, 
\begin{equation}
n=2h_{1}=2h_{2}\,,\label{eqn:higgs_2g_helicity_eqn}
\end{equation}
which immediately sets
\begin{align}
\mathcal{M}\left(h; g^{+}g^{-}\right)=0 \,.
\end{align}
The only relevant amplitude to consider is then
$\mathcal{M}\left(h; ++\right)$ (with
$\mathcal{M}\left(h; --\right)$ determined by a parity transformation).
Then $n=2$,
and since the amplitude has mass dimension $1$,
$\ell=1$ and 
\begin{align}
\mathcal{M}\parn{h; g^{a+}(p_1)g^{b+}(p_2)} &= \delta^{ab} \,
\left[12\right]^2 f_{-1}\left(m^2,\Lambda^2\right)
=\delta^{ab} \, \frac1\Lambda\, \left[12\right]^2\,\, \tilde f\left(\frac{m^2}{\Lambda^2}\right) \,,
\end{align}
where $\tilde f$ is  dimensionless.    
Note that~\eqref{higgs_2g_helicity_eqn}, combined with the mass dimension of the amplitude,  gives a selection rule relating the sum of the gluon helicities to the dimension
of the coupling which generates the amplitude (see also~\cite{Azatov:2016sqh}).
Specifically, here
\begin{align}
	\left\vert h_1+h_2\right\vert = l+1\,,
\end{align}
with $l=1$.

At tree-level, the function $\tilde f$ can be written as a power series in $m^2$.
No negative power of $m^2$ can appear, since the amplitude must vanish for $m \to 0$.
The amplitude is therefore given by,
\begin{align}
\label{eqn:h_2g_final_amp}
\mathcal{M}\parn{h; g^{a+}(p_1)g^{b+}(p_2)} &=\delta^{ab} \,
\frac{\left[12\right]^2}{\Lambda}\,
\sum_{n=0}^{\infty} c_n \left(\frac{m^2}{\Lambda^2}\right)^n \equiv \delta^{ab} \, \frac{c^{hgg}_{5}}{\Lambda} \, \sqb{12}^{2}  \,,
\end{align}
where we rescaled the infinite series into the coefficient $c^{hgg}_{5}$.
This is indeed the most general three-point amplitude for one massive scalar
and two massless
vectors~\cite{Conde:2016vxs,Arkani-Hamed:2017jhn}.

We can now make contact with the EFT calculation.
The lowest order operators mediating scalar decay to two spin-1 particles
are dimension-5. In a CP-conserving theory, there is only a single such operator,
namely~\eqref{smhiggs},
in accord with the single real coefficient $c^{hgg}_{5}$  at this order.
Operators of higher dimension which contribute to the amplitude  still
contain two powers of the field-strength $G$, but an even power of derivatives.
Since we consider a purely gluonic theory with no quarks, 
the EOM is $D^\mu G_{\mu\nu}=0$.
Using this and integration by parts,
there is only a single  independent operator at each order in $\Lambda$,
with the derivatives acting on $h$ and giving powers of $m^2$.
In this case, this series merely gives a rescaling of $c^{hgg}_{5}$.

In associating the amplitude~\eqref{h_2g_final_amp} with the operator~\eqref{smhiggs}
we have assumed a scalar $h$ and CP invariance.
A pseudo-scalar would couple to the operator $G\tilde G$, with an
identical result for the amplitude.
In fact, a better way to organize the theory is by grouping the scalar and pseudo-scalar into a complex field $\phi$, with the Lagrangian~\cite{Dixon:2004za}
\begin{equation}
	\phi \,\GSD^2 + \phi^{\dagger}\, \GASD^2 \,,
\end{equation}
where the selfdual and anti-selfdual field strengths are defined as 
\begin{align}
	G^{\mu\nu}_{\mathrm{SD}} = \frac{1}{2}\parn{G^{\mu\nu} + \tilde{G}^{\mu\nu}}\,, \quad G^{\mu\nu}_{\mathrm{ASD}} = \frac{1}{2}\parn{G^{\mu\nu}- \tilde{G}^{\mu\nu}} \,, \quad \tilde{G}^{\mu\nu} = \frac{i}{2}\epsilon^{\mu\nu\rho\sigma}G_{\rho\sigma}\,.
\end{align}
From the EFT point of view, the amplitude we calculated corresponds to the operator $\phi \,\GSD^2$, since $\GSD$ generates positive helicity gluons,
while $\GASD$ generates negative helicity gluons~\cite{Dixon:2004za,Witten:2003nn}.
The amplitude mediated by $\phi^{\dagger} \GASD^2$ can be obtained from this amplitude by reversing the gluon helicities, and switching angle and square brackets as explained in~\cite{Dixon:2004za}. 
The scalar (pseudo scalar) amplitude is then obtained as the sum (difference) of the $\phi$ and $\phi^{\dagger}$ amplitudes.
The $\phi$ and $\phi^\dagger$ 2-gluon decay amplitudes are identical, but, allowing for CP-violation, the coefficient $c^{hgg}_{5}$ may be complex. 
In any case, the 2-gluon decay cannot distinguish between a scalar and a pseudo-scalar.
This would require at least 4 independent momenta, namely a 4-gluon final state.

\subsection{The scalar plus 3 gluon amplitudes ~${\cal M}(h;ggg)$}

The scalar plus three-gluon amplitude
can be written in terms of
the three helicity brackets and three Lorentz invariants, 
\begin{align}
	\mathcal{M}\left(h; g^{a, \, h_1}(p_1)
	g^{b, \, h_2}(p_2)  g^{c, \, h_3}(p_3)
	\right) &= C^{abc}\,
	\left[12\right]^{n_{12}}\left[13\right]^{n_{13}}\left[23\right]^{n_{23}}
	\,
	f_{-\ell}\left(s_{12},s_{23},s_{13};\Lambda\right)\,,
	\label{eqn:higgs_3g_ansatz}
\end{align}
where $-\ell$ denotes the mass dimension of the function $f_{-\ell}$,
and the color factor $C^{abc}$ is either $f^{abc}$ or $d^{abc}$.
Since the amplitude has mass dimension zero,
\begin{align}
	\ell= n_{12}+n_{13}+n_{23}\,.
\end{align}
Little group scaling determines the powers $n_{ij}$ in terms of the gluon
helicities\footnote{Choosing to work with angle brackets instead,
with powers $m_{12}$,
$m_{13}$, $m_{23}$ would give	
$\ell =-h_{1}-h_{2}-h_{3}$ and 
$m_{12} =\ell -2h_{3}$ etc.},
\beq\label{eqn:higgs_3g_helicity_eqns}
	n_{12} + n_{13}  = 2h_{1} \,,~~~
	n_{12} + n_{23}  =	2h_{2} \,,~~~
	n_{13} + n_{23}  =	2h_{3} \,.
\eeq
so
\begin{align}\label{eqn:higgs_3g_mass_dimension_coupling}
	&n_{12} =\ell - 2h_{3} \eqCS
	n_{13}  =\ell - 2h_{2} \eqCS
	n_{23}  =\ell - 2h_{1} \,.
\end{align}
As before, \eqref{higgs_3g_mass_dimension_coupling} relates the sum of helicities to the dimension of the coupling which generates the amplitude,
\begin{align}
	\ell &= h_1+h_2+h_3\,,
\end{align}
so $\ell$ is odd, as in the 2-gluon case, and we immediately see that
the amplitude must be generated by a higher dimension operator: the only invariants associated with bosons have mass-dimension 2, so at least one power of $\Lambda$ is required.

At tree-level, since we consider only operators with a single $h$,
the function $f_{-\ell}$ is simply a power series in the $s_{ij}$'s, and the only possible poles are single poles in the $s_{ij}$'s coming from gluon propagators.
This part of the amplitude factorizes as the product of two 3-particle amplitudes,
${\cal M}(hgg)$ of~\eqref{h_2g_final_amp},
and ${\cal M}(ggg)$ which we recall below.
The all-plus amplitude is thus of the form,
\begin{align}
\label{eqn:higgs_3g_amp_initial}
\mathcal{M}\parn{h; g^{a+}\parn{p_{1}} g^{b+}\parn{p_{2}} g^{c+}\parn{p_{3}}} &= \frac{1}{\Lambda} \, \left[12\right]\left[23\right]\left[13\right] \, C^{abc} \times \nn \\
&\times\bigg[\sum_{\substack{n,k,l=0 }}^{\infty} \frac{a_{n,k,l}}{\Lambda^{2(n+k+l+1)}} \, s_{12}^{n} s_{13}^{k} s_{23}^l + \textrm{factorizable} \bigg] \,, 
\end{align}
subject to the constraint $s_{12}+s_{23}+s_{13}=m^2$.
Here $a_{n,k,l}$ are dimensionless constants which, for $C^{abc} = f^{abc} \: \parn{d^{abc}}$, are
completely symmetric (antisymmetric) in the indices $n,k,l$.

Similarly, for $h_{1}=h_{2}=-h_{3}=1$, \eqref{higgs_3g_mass_dimension_coupling} yields the solution
\begin{align}\label{eqn:higgs_3g_plus_plus_minus_amp_initial}
	\mathcal{M}\parn{h; g^{a+}\parn{p_{1}} g^{b+}\parn{p_{2}} g^{c-}\parn{p_{3}}} &= \frac{\left[12\right]^{3}}{\left[13\right]\left[23\right]}\frac{C^{abc}}{\Lambda} \times \nn \\
	&\times\bigg[\sum^{\infty}_{\substack{n,k,l=0 }} \frac{b_{n,k,l}}{\Lambda^{2\parn{n+k+l}}} \, s^{n}_{12} s^{k}_{13} s^{l}_{23} + \textrm{factorizable}\bigg]\,,
\end{align}
where for $C^{abc} = f^{abc}$ we have $b_{n,k,l}=b_{n,l,k}$, while
for $C^{abc} = d^{abc}$,   $b_{n,k,l}~=-b_{n,l,k}$.

\subsubsection{Three-gluon vertices}
To calculate the factorizable parts of the Higgs plus 3-gluon amplitudes,
we  need three-gluon vertices.
These can be obtained as three-gluon amplitudes with complex momenta,
for which either $[ij]$ or $\langle ij\rangle$ vanish, such that
$s_{ij}=0$. 
Based on little group considerations and the mass dimension of the
3-point amplitude, the $++-$ and $+++$ tree-level amplitudes must be of the
form (see~\eg~\cite{Elvang:Book15})
\begin{subequations}\label{eqn:gluing_amp_3g}
	\begin{align}		
	  \mathcal{M}_{3g}\parn{1^{a+} 2^{b+} 3^{c-}} &= f^{abc} \, g_{s} \, \frac{\sqb{12}^{3}}{\sqb{13}\sqb{23}}
          \label{eqn:gluing_amp_3g_tr_QCD}\,,\\
	\mathcal{M}_{3g}\parn{1^{a+} 2^{b+} 3^{c+}} &= f^{abc} \, c^{ggg}_{6} \, \frac{\sqb{12}\sqb{13}\sqb{23}}{\Lambda^2}  \label{eqn:gluing_amp_3g_tr_GGG}\,, 
        \end{align}
\end{subequations}
where $c^{ggg}_{6}$ is a dimensionless coefficient, and $g_{s}$ is 
the strong coupling (up to a convention dependent numerical coefficient).
The function $f_{-\ell}(s_{ij})=1$ in this case, 
since $s_{ij}=0$. These
vertices therefore receive no further contributions. 

Again, we see that vertices of different net helicities arise from couplings of different dimensions.
The vertex of~\eqref{gluing_amp_3g_tr_QCD} is the QCD vertex, while 
the vertex of~\eqref{gluing_amp_3g_tr_GGG}, comes from the dimension--6 operator
$\Tr\parn{G^{3}}$.
As pointed out in~\cite{Azatov:2016sqh}, this explains why this operator does not
affect dijet production at leading
order~\cite{Simmons:1989zs,Dixon:1993xd}~\footnote{At higher order, this is no longer
  the case. Thus for example the helicity structure of the tree-level 5-gluon amplitude,
  or one-loop 4-gluon amplitudes, generated by this operator is not
  orthogonal to QCD~\cite{Dixon:1993xd}.}.

Note that $c^{ggg}_{6}$ is in principle complex, corresponding to the two operators $\Tr\parn{G^{3}}$ and $\Tr\parn{G^{2}\tilde{G}}$.

\subsubsection{The $h;+++$ amplitude}

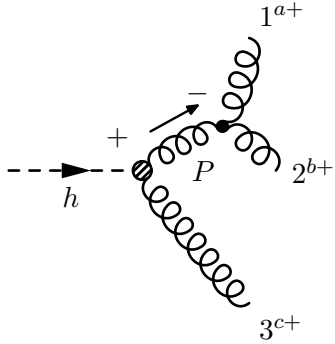
\begin{figure}
	\centering
	\begin{fmffile}{h_3g_QCD_factorization_ppp} 
		\fmfcmd{style_def marrow expr p = drawarrow subpath (1, 0.3) of p shifted 13 up
			withpen pencircle scaled 0.9; enddef;}
		\fmfcmd{style_def marrowA expr p = drawarrow subpath (3/4, 1/4) of p shifted 6 up
			withpen pencircle scaled 0.4; label.top(btex $p_{2}$ etex, point 1 of p
			shifted 15 up); enddef;}
		\begin{fmfgraph*}(100,100)         
			\fmfleftn{i}{1} 
			\fmfrightn{o}{5}
			\fmfdotn{v}{2} 
			\fmf{phantom, tension=2}{i1,v1,o3}
			\fmffreeze
			\fmf{scalar, label=$h$, tension=2}{i1,v1}
			\fmf{gluon}{v1,o1}
			\fmf{gluon}{o3,v2,o5}
			\fmf{gluon, label=$P$, label.side=left}{v2,v1} 
			\fmffreeze
			\fmf{marrow}{v2,v1} 
			\fmfblob{0.07w}{v1}
			\fmflabel{$1^{a+}$}{o5}
			\fmflabel{$2^{b+}$}{o3}
			\fmflabel{$3^{c+}$}{o1}
			\fmfv{l=$+$,l.a=120,l.d=.10w}{v1}
			\fmfv{l=$-$,l.a=120,l.d=.10w}{v2}
		\end{fmfgraph*} 
	\end{fmffile}
	\bigskip
	\caption{\label{fig:h_3g_QCD_factorization_ppp}Factorization of the $+++$ amplitude on $s_{12}$. The direction of the arrow indicates the direction of momentum flow. }
\end{figure}
%
We begin by computing the factorizable part of the amplitude of~\eqref{higgs_3g_amp_initial}.
Consider the $s_{12}$ pole. Because we  pulled out a factor of $\sqb{12}$ we must set $\anb{12}$ to zero. 
Only the $+++$ 3-gluon amplitude can contribute in this case
(see~\figref{h_3g_QCD_factorization_ppp}),
and we get\footnote{Note that  the momentum in the $h\to 2g$ vertex is $-P$, because
  we take all the momenta to be incoming.
  In the notation of~\cite{Schwartz:Book2014}, spinors with $-P$ are related to
  spinors with $P$ by a factor of $i$, \eg~$\ket{-P} = i\ket{P}$ etc. }
\begin{align}
	\mathcal{M}\parn{h; g^{a+}\parn{p_{1}} g^{b+}\parn{p_{2}} g^{c+}\parn{p_{3}}} & =  \mathcal{M}_{3g}\parn{1^{a+}2^{b+}P^{-}} \, \frac{-i}{s_{12}} \, \mathcal{M}_{hgg}\parn{3^{+}\parn{-P}^{+}} \nn \\
		& = f^{abc} \, \tilde{g}_{s} \, \frac{\sqb{12}^{3}}{\sqb{1P}\sqb{2P}} \, \frac{i}{s_{12}} \, \frac{c^{hgg}_{5}}{\Lambda} \, \sqb{P3}^{2} \nn \\
		& = f^{abc} \, \tilde{g}_{s}\, \frac{\sqb{12}^{3}}{\sqb{1P}\anb{P3}\sqb{2P}\anb{P3}} \, \frac{i}{s_{12}} \, \frac{ c^{hgg}_{5}}{\Lambda} \, \parn{s_{23}+s_{13}}^{2} \nn\\
		& = -\frac{i\,c_{5}^{hgg}\,g_{s}}{\Lambda}\,\frac{m^{4}\,\sqb{12}\sqb{13}\sqb{23}}{s_{12}s_{13}s_{23}}\,f^{abc} \,,
\end{align}
where we have used $\lim_{\anb{12}\to 0} \parn{s_{23}+s_{13}} = m^{2}$ and $P = -p_1 -p_2$.
Since the amplitude is completely symmetric under $i\leftrightarrow j$ we need not consider factorizations under $s_{23}\,, s_{13}$.

Substituting this in \eqref{higgs_3g_amp_initial} we then have for the
$+++$ amplitude,
\begin{align}\label{eqn:h_3g_+++_full_constrained_amplitude}
	\mathcal{M}\parn{h; g^{a+}\parn{p_{1}} g^{b+}\parn{p_{2}} g^{c+}\parn{p_{3}}}  &= \frac{\sqb{12}\sqb{13}\sqb{23}}{\Lambda} \, 
		\bigg[ 
			-i\,f^{abc}\,g_{s}\,c_{5}^{hgg}\,\frac{m^{4}}{s_{12}s_{13}s_{23}}  \nn\\
			&+\sum^{\infty}_{\substack{n,k,l=0}} \frac{a_{n,k,l} \, f^{abc} + \alpha_{n,k,l} \, d^{abc}}{\Lambda^{2\parn{n+k+l+1}}} s^{n}_{12} s^{k}_{13} s^{l}_{23}
		\bigg]\,.
\end{align}
Expanding the series~\eqref{h_3g_+++_full_constrained_amplitude} up to $n+k+l \le 4$, \ie, dimension 13, we get
\begin{align}\label{eqn:h_3g_+++_full_constrained_amplitude_up_to_dim_13}
	&\mathcal{M}\parn{h; g^{a+}\parn{p_{1}} g^{b+}\parn{p_{2}} g^{c+}\parn{p_{3}}}  = \frac{\sqb{12}\sqb{13}\sqb{23}}{\Lambda} \, 
	\bigg[ 
	f^{abc} \, \bigg( 
	-i\,\frac{m^{4}\,g_{s}\,c_{5}^{hgg}}{s_{12}s_{13}s_{23}} + \frac{a_{0,0,0}}{\Lambda^{2}} + \frac{a_{1,0,0}\parn{s_{12}+s_{13}+s_{23}}}{\Lambda^{4}} \nn \\
	&+\frac{1}{\Lambda^{6}}\parn{a_{2,0,0}\parn{s^{2}_{12} + s^{2}_{13} + s^{2}_{23}} + a_{1,1,0} \parn{s_{12}s_{13} + s_{12}s_{23} + s_{13}s_{23}}}
	+ \frac{1}{\Lambda^{8}}\bigg( a_{1,1,1}s_{12} s_{13} s_{23} \nn \\ &+ a_{0,0,3}\parn{s_{12}^3+s_{13}^3+s_{23}^3}   
	+a_{0,1,2}\parn{s_{13} s_{12}^2+s_{23} s_{12}^2+s_{13}^2 s_{12}+s_{23}^2 s_{12}+s_{13} s_{23}^2+s_{13}^2 s_{23}}  
	\bigg)
	\bigg)	\nn \\
	&+\frac{d^{abc}}{\Lambda^{8}} \,  \left(s_{12}-s_{13}\right) \left(s_{12}-s_{23}\right) \left(s_{13}-s_{23}\right)\alpha_{0,1,2}
	\bigg]\,.
\end{align}
This result may be simplified by using momentum conservation, which implies $s_{12}+s_{23}+s_{23} = m^{2}$. 
We may also trade $s^{2}_{12}+s^{2}_{13}+s^{2}_{23}$ for
${\parn{s_{12}+s_{23}+s_{13}}^{2}} - 2s_{13}s_{12}...$, and similarly for
the cubic polynomial.
The final result for the EFT amplitude up to dimension 13 is
\begin{align}\label{eqn:higgs_3g_all_plus_final_result}
	&\mathcal{M}\parn{h; g^{a+}\parn{p_{1}} g^{b+}\parn{p_{2}} g^{c+}\parn{p_{3}}}  = \frac{\sqb{12}\sqb{13}\sqb{23}}{\Lambda} \, 
	\bigg[
		f^{abc} \, 
		\bigg(
			-i\,\frac{m^{4}\,g_{s}\,c_{5}^{hgg}}{s_{12}s_{13}s_{23}} + \frac{a_{7}}{\Lambda^{2}} \nn \\
			&+ \frac{a_{11}}{\Lambda^{6}}\parn{s_{12}s_{23}+s_{13}s_{23}+s_{12}s_{13}} + \frac{a_{13}}{\Lambda^{8}}s_{12}s_{13}s_{23}
		\bigg) \nn \\
		&+ d^{abc} \, \frac{a^{\prime}_{13}}{\Lambda^{8}} \,  \left(s_{12}-s_{13}\right) \left(s_{12}-s_{23}\right) \left(s_{13}-s_{23}\right)
	\bigg]\,,
\end{align}
where we defined
\begin{subequations}
	\begin{align}
	a_{7}  &\equiv  a_{0,0,0}+a_{1,0,0}\frac{m^{2}}{\Lambda^{2}}+a_{2,0,0}\frac{m^{4}}{\Lambda^{4}}+a_{0,0,3}\frac{m^{6}}{\Lambda^{6}} \,,\\
	a_{11} &\equiv  a_{1,1,0}-2a_{2,0,0}+\left(a_{0,1,2}-3a_{0,0,3}\right)\frac{m^{2}}{\Lambda^{2}} \,,\\ 
	a_{13} &\equiv  a_{1,1,1}-3a_{0,1,2}+3a_{0,0,3}\,,
	\end{align}
\end{subequations}
and $a^{\prime}_{13} \equiv \alpha_{0,1,2} $. 
The amplitude for a massless $h$ is identical. We simply set $m^2=0$ in intermediate stages above, with the end result unchanged.

\subsubsection{The $h;++-$ amplitude}
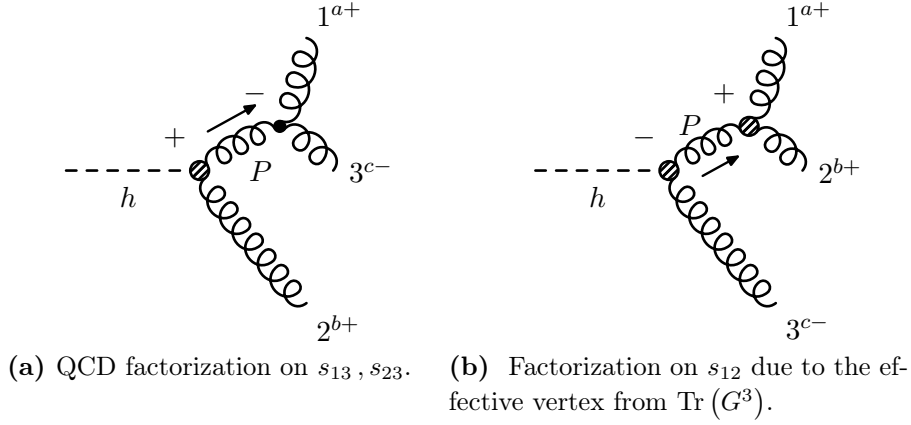
\begin{figure}
	\centering
	\begin{subfigure}[t]{0.4\textwidth}
		\centering
		\begin{fmffile}{h_3g_dim6_factorization_ppm} 
			\fmfcmd{style_def marrow expr p = drawarrow subpath (1, 0.3) of p shifted 13 up
				withpen pencircle scaled 0.9; enddef;}
			\fmfcmd{style_def marrowA expr p = drawarrow subpath (3/4, 1/4) of p shifted 6 up
				withpen pencircle scaled 0.4; label.top(btex $p_{2}$ etex, point 1 of p
				shifted 15 up); enddef;}
			\begin{fmfgraph*}(100,100)         
				\fmfleftn{i}{1} 
				\fmfrightn{o}{5}
				\fmfdotn{v}{2} 
				\fmf{phantom, tension=2}{i1,v1,o3}
				\fmffreeze
				\fmf{dashes, label=$h$, tension=2}{i1,v1}
				\fmf{gluon}{v1,o1}
				\fmf{gluon}{o3,v2,o5}
				\fmf{gluon, label=$P$, label.side=left}{v2,v1} 
				\fmffreeze
				\fmf{marrow}{v2,v1} 
				\fmfblob{0.07w}{v1}
				\fmflabel{$1^{a+}$}{o5}
				\fmflabel{$3^{c-}$}{o3}
				\fmflabel{$2^{b+}$}{o1}
				\fmfv{l=$+$,l.a=120,l.d=.10w}{v1}
				\fmfv{l=$-$,l.a=120,l.d=.10w}{v2}
			\end{fmfgraph*} 
		\end{fmffile}
		\bigskip
		\caption{\label{fig:h_3g_plus_plus_minus_QCD_factorization}QCD factorization on $s_{13}\,, s_{23}$.}
	\end{subfigure}
	\begin{subfigure}[t]{0.4\textwidth}
		\centering
		\begin{fmffile}{h_3g_QCD_factorization_ppm} 
			\fmfcmd{style_def marrow expr p = drawarrow subpath (0.6, 0.05) of p shifted 9 down
			withpen pencircle scaled 0.9; enddef;}
			\fmfcmd{style_def marrowA expr p = drawarrow subpath (3/4, 1/4) of p shifted 6 up
				withpen pencircle scaled 0.4; label.top(btex $p_{2}$ etex, point 1 of p
				shifted 15 up); enddef;}
			\begin{fmfgraph*}(100,100)         
				\fmfleftn{i}{1} 
				\fmfrightn{o}{5}
				\fmfdotn{v}{2} 
				\fmf{phantom, tension=2}{i1,v1,o3}
				\fmffreeze
				\fmf{dashes, label=$h$, tension=2}{i1,v1}
				\fmf{gluon}{v1,o1}
				\fmf{gluon}{o3,v2,o5}
				\fmf{gluon, label=$P$, label.side=right}{v2,v1} 
				\fmffreeze
				\fmf{marrow}{v2,v1} 
				\fmfblob{0.07w}{v2}
				\fmfblob{0.07w}{v1}
				\fmflabel{$1^{a+}$}{o5}
				\fmflabel{$2^{b+}$}{o3}
				\fmflabel{$3^{c-}$}{o1}
				\fmfv{l=$-$,l.a=120,l.d=.10w}{v1}
				\fmfv{l=$+$,l.a=120,l.d=.10w}{v2}
			\end{fmfgraph*} 
		\end{fmffile}
		\bigskip
		\caption{\label{fig:h_3g_plus_plus_minus_Dim6_factorization} Factorization on $s_{12}$ due to the effective vertex from $\Tr\parn{G^{3}}$.}
	\end{subfigure}
	\caption{Possible factorizations of the $++-$ amplitude.}
\end{figure}

Again, we begin by considering the factorizable part. 
In this case, both the $+++$ and $+--$  3-gluon amplitudes can contribute.
The $++-$ contribution from the $s_{13}$ pole (see~\figref{h_3g_plus_plus_minus_QCD_factorization}) is,
\begin{align}
	\mathcal{M}\parn{h; g^{a+}\parn{p_{1}} g^{b+}\parn{p_{2}} g^{c-}\parn{p_{3}}} &= \mathcal{M}_{3g}\parn{1^{a+}3^{c-}P^{-}}\,\frac{-i}{s_{13}}\,\mathcal{M}_{hgg}\parn{2^{b+}\left(-P\right)^{+}} \nn\\
		&=\frac{ig_{s}\,c_{5}^{hgg}}{\Lambda} \, \frac{\left[12\right]^{3}}{\left[13\right]\left[23\right]} \, f^{abc} \,.
\end{align}
The $s_{12}$ pole is shown in~\figref{h_3g_plus_plus_minus_Dim6_factorization} and gives,
\begin{align}
	\mathcal{M}\parn{h; g^{a+}\parn{p_{1}} g^{b+}\parn{p_{2}} g^{c-}\parn{p_{3}}} &= \mathcal{M}_{3g}\parn{1^{a+}2^{b+}P^{+}}\,\frac{-i}{s_{12}}\,\mathcal{M}_{hgg}\parn{3^{c-}\left(-P\right)^{-}} \nn\\
		&=-\frac{ic_{5}^{hgg}\,c_{6}^{ggg}}{\Lambda^{3}}\,\frac{\sqb{12}^{3}}{\left[13\right]\left[23\right]}\,\frac{s_{23}s_{13}}{s_{12}} \, f^{abc}\,.
\end{align}
%
%
%
The full EFT amplitude is then given by
\begin{align}\label{eqn:h_3g_++-_full_constrained_amplitude}
	\mathcal{M}\parn{h; g^{a+}\parn{p_{1}} g^{b+}\parn{p_{2}} g^{c-}\parn{p_{3}}}  &= \frac{\sqb{12}^3}{\sqb{13}\sqb{23}} \, \frac{1}{\Lambda} 
	\bigg[ 
			f^{abc} \, \bigg( ig_{s}\,c_{5}^{hgg} -\frac{ic_{5}^{hgg}\,c_{6}^{ggg}}{\Lambda^{2}}\,\frac{s_{23}s_{13}}{s_{12}} \bigg) \nn\\ 
			&+\sum^{\infty}_{\substack{n,k,l=0}} \frac{b_{n,k,l} \, f^{abc} + \beta_{n,k,l} \, d^{abc}}{\Lambda^{2\parn{n+k+l}}} s^{n}_{12} s^{k}_{13} s^{l}_{23}
	\bigg]\,,
\end{align}
where $b_{n,k,l} = b_{n,l,k}$ and $\beta_{n,k,l} = -\beta_{n,l,k}$. 
Computing the series in~\eqref{h_3g_++-_full_constrained_amplitude} up to 
dimension 13, and discarding redundant terms which are related by powers of ${m^2}/{\Lambda^2}$ as before, we get
\begin{align}\label{eqn:h_3g_++-_final_result}
	&\mathcal{M}\parn{h; g^{a+}\parn{p_{1}} g^{b+}\parn{p_{2}} g^{c-}\parn{p_{3}}}   =
		 \frac{\sqb{12}^3}{\sqb{13}\sqb{23}} \, \frac{1}{\Lambda} 
		\bigg[ 
			f^{abc} \,
			\bigg(
				ig_{s}\,c_{5}^{hgg} -\frac{ic_{5}^{hgg}\,c_{6}^{ggg}}{\Lambda^{2}}\,\frac{s_{23}s_{13}}{s_{12}} + \frac{b_{9}}{\Lambda^{4}}\,s_{13} s_{23} \nn\\ &+ \frac{b_{11}}{\Lambda ^6} \, s_{12} s_{13} s_{23}
				+ \frac{b_{13}}{\Lambda ^8} \, s_{13}^{2} s_{23}^{2}  +\frac{b^{\prime}_{13}}{\Lambda ^8}\,s_{13} s_{23} s_{12}^2 
			\bigg) 
			+ d^{abc} \, s_{13}s_{23}\parn{s_{13}-s_{23}} \, \bigg(\frac{b^{\prime}_{11}}{\Lambda ^6} + \frac{b^{\prime\prime}_{13}}{\Lambda ^8}\, s_{12}\bigg)  
		\bigg]\,,
\end{align}
with an identical result for a massless $h$.

Finally, note that the scalar plus photon amplitudes,
$h\gamma\gamma$ and $h\gamma\gamma\gamma$, can be obtained from our results above
by setting $f^{abc}=0$ and omitting $\delta^{ab}$ and $d^{abc}$.

\subsubsection{Inferring the EFT Lagrangian}
\begin{table}[t]
	\centering
	\caption{Operators contributing to the $h\to 3g$ amplitude.
		The number of independent operators of each type
		and color structure appears in brackets.}
	\label{tab:EFT_h_3g_operator_list}
	\begin{tabular}{|c|c|c|}
		\hline 
		\multirow{2}{*}{Mass dimension} & \multicolumn{2}{c|}{Operators}\\
		\cline{2-3} 
		& $\mathcal{M}\parn{+++}$ & $\mathcal{M}
		\parn{++-}$\\
		\hline 
		\hline 
		5 & --- & --- \\
		\hline 
		7 & $h \, \GSD^{3}$~$[1, f^{abc}]$&  ---\\
		\hline 
		9 & --- & $\mathcal{D}^2 \, \GSD^{2} \, \GASD \, h
		$~$[1, f^{abc}]$ \\
		\hline 
		11 & $\mathcal{D}^4 \, \GSD^3 \, h$~$[1, f^{abc}]$
		& $\, \mathcal{D}^4 \, \GSD^{2} \, \GASD \, h$~$[1, f^{abc};
		1, d^{abc}]$\\
		\hline 
		13 &
		$\, \mathcal{D}^6 \, \GSD^3 \, h$~$[1, f^{abc}; 1, d^{abc}]$ &
		$ \, \mathcal{D}^6 \, \GSD^{2} \, \GASD\, h$~$[2, f^{abc};
		1, d^{abc}]$\\
		\hline 
	\end{tabular}
\end{table}
\eqsref{higgs_3g_all_plus_final_result}{h_3g_++-_final_result} represent
the EFT contribution to the $h;+++$ and $h;++-$ amplitudes respectively,
up to dimension 13. 
Apart from the couplings $g_s$, $c^{ggg}_{6}$, $c^{hgg}_{5}$,
associated with the three relevant three-particle amplitudes,
the $h;+++$ amplitude depends on four distinct kinematical structures,
with four independent coefficients:
$a_{7}$, $a_{11}$, $a_{13}$ and $a^{\prime}_{13}$.
The $h;++-$ amplitude has six distinct kinematical structures, with six independent coefficients: $b_{9}$, $b_{11}$, $b^{\prime}_{11}$, $b_{13}$ $b^{\prime}_{13}$
and $b^{\prime\prime}_{13}$. 
From the EFT Lagrangian point of view, these are associated
with the coefficients of operators with a single $h$, three powers of the
field strength, and some number of covariant derivatives.
Recalling  that positive-helicity (negative-helicity)
gluons correspond to
$\GSD^{\mu\nu}$ ($\GASD^{\mu\nu}$),
we can infer the schematic form of the operators, and the number of
independent operators at each dimension.
We display these in Table~\ref{tab:EFT_h_3g_operator_list}, where we divided
the operators according to whether they contribute to the
$+++$ or $++-$ amplitudes. The number of independent operators of each type,
with color indices contracted with $f^{abc}$ or $d^{abc}$,
is indicated in parenthesis.
As a check of our results, we used the Mathematica notebook 
of~\cite{Henning:2015alf} to derive the scalar plus 3 gluon operators,
and verified that the results agree.

\subsection{The two scalar, two gluon amplitudes ~${\cal M}(hh;gg)$}
\begin{figure}[t]
	\centering
	\begin{subfigure}[t]{0.4\textwidth}
		\centering
		\begin{fmffile}{2h_2g_factorization_3h_vtx} 
			\fmfcmd{style_def marrow expr p = drawarrow subpath (0.85, 0.15) of p shifted 8 left
				withpen pencircle scaled 0.9; enddef;}
			\fmfcmd{style_def marrowA expr p = drawarrow subpath (3/4, 1/4) of p shifted 6 up
				withpen pencircle scaled 0.4; label.top(btex $p_{2}$ etex, point 1 of p
				shifted 15 up); enddef;}
			\begin{fmfgraph*}(100,100)         
				\fmfleftn{i}{2} 
				\fmfrightn{o}{2}
				\fmfdotn{v}{2} 
				\fmf{gluon}{i1,v1,i2}
				\fmf{dashes}{v1,v2}
				\fmf{dashes}{o1,v2,o2}
				\fmffreeze 
				\fmfblob{0.07w}{v1}
				\fmflabel{$1^{a+}$}{i1}
				\fmflabel{$2^{b+}$}{i2}
				\fmflabel{$3$}{o1}
				\fmflabel{$4$}{o2}
			\end{fmfgraph*} 
		\end{fmffile}
		\bigskip
		\caption{\label{fig:2h_2g_factorization_dim_5_three_point_higgs}Factorization on $h \to 2g$ and the $3h$ vertex.}
	\end{subfigure}
	\begin{subfigure}[t]{0.4\textwidth}
		\centering
		\begin{fmffile}{2h_2g_factorization} 
			\fmfcmd{style_def marrow expr p = drawarrow subpath (0.85, 0.15) of p shifted 8 left
				withpen pencircle scaled 0.9; label.top(btex $P$ etex, point 0.7 of p
								shifted 15 right); enddef;}
			\begin{fmfgraph*}(100,100)         
				\fmfleftn{i}{2} 
				\fmfrightn{o}{2}
				\fmfdotn{v}{2} 
				\fmf{gluon}{i1,v1,v2,i2}
				\fmf{dashes}{v1,o1}
				\fmf{dashes}{v2,o2}
				\fmffreeze
				\fmf{marrow}{v2,v1} 
				\fmfblob{0.07w}{v1}
				\fmfblob{0.07w}{v2}
				\fmflabel{$1^{a-}$}{i1}
				\fmflabel{$2^{b+}$}{i2}
				\fmflabel{$3$}{o1}
				\fmflabel{$4$}{o2}
				\fmffreeze
				\fmfv{l=$-$,l.a=0,l.d=.10w}{v1}
				\fmffreeze
				\fmfv{l=$+$,l.a=0,l.d=.10w}{v2}
			\end{fmfgraph*} 
		\end{fmffile}
		\bigskip
		\caption{\label{fig:2h_2g_factorization_dim_5}Factorization on the dimension~5 amplitudes~$h \to 2g$.
                  Note that there is a crossed diagram with $3\leftrightarrow 4$. }
	\end{subfigure}
	\caption{Possible factorizations of the $2h \to 2g$ amplitude.}
\end{figure}
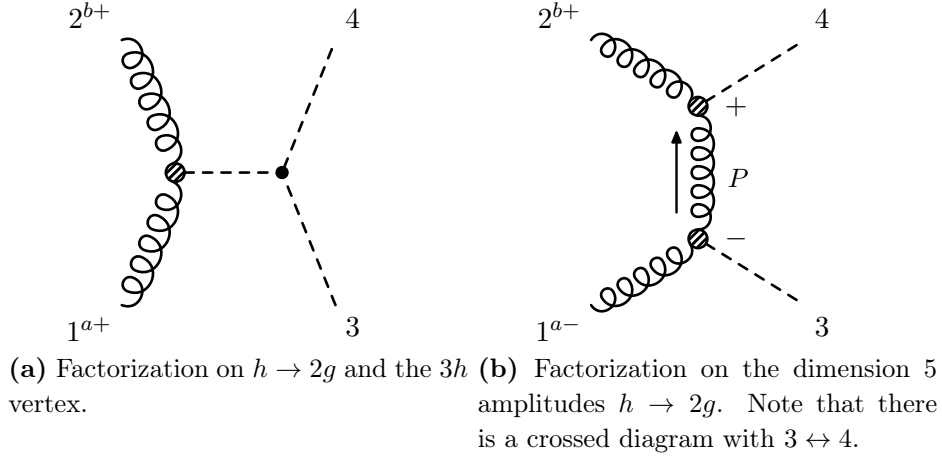
%
To calculate the two-$h$, two-gluon amplitude, it is convenient to
keep the symmetry under both $h$ exchange and gluon exchange manifest.
Therefore, we will not use momentum  conservation to eliminate any of
the external momenta.
Let us label the massless momenta by $p_{1}\,, p_{2}$, and the massive momenta by $p_{3}$ and $p_{4}$. 
There are two independent structures which carry little group weights in
this case, namely, 
\begin{equation}
  \sqb{12}\,~~{\rm and}~\braKet{1}{p_{3}}{2}\,.
  \label{eqn:di_higgs_2g_invariants}
\end{equation}
Thus the $2h;2g$ amplitude is of the form, 
\begin{equation}
\mathcal{M}\parn{2h; g^{a, \, h_{1}}\parn{p_{1}} g^{b, \, h_{2}}\parn{p_{2}}} = \delta^{ab} \, \left[12\right]^{n_{12}}\braKet{1}{p_{3}}{2}^{k_{12}}f_{-\ell}\,,
\end{equation}
where $n_{12}\,,\:k_{12}$ are integers and $f_{-\ell}$ is an analytic function of mass dimension $-\ell$. 
Requiring the correct helicity weights and mass dimension yields the equations,
\begin{align}
	&-2h_{1} = k_{12} - n_{12}\,, \quad  -2h_{2}  =-k_{12}-n_{12}    \,, \quad \ell =n_{12}+2k_{12} \,,
\end{align}
so
\begin{align}\label{eqn:2h_2g_helicity_mass_eqn_sol}
& n_{12}=h_{1}+h_{2} \eqCS k_{12}=h_{2}-h_{1} \eqCS \ell=3h_{2}-h_{1}\,.
\end{align}
Plugging these into~\eqref{2h_2g_helicity_mass_eqn_sol} we get the amplitudes
\begin{subequations}\label{eqn:2h_2g_amplitudes_initial}
	\begin{align}
		\mathcal{M}\parn{2h; g^{a+}\parn{p_{1}} g^{b+}\parn{p_{2}}} & =\delta^{ab} \, \left[12\right]^{2}f_{-2}\,,\\
		\mathcal{M}\parn{2h; g^{a-}\parn{p_{1}} g^{b+}\parn{p_{2}}} & = \delta^{ab} \, \braKet 1{p_{3}}2^{2}\tilde{f}_{-4}=
 \delta^{ab} \, \frac14\braKet 1{p_{3}-p_4}2^{2}\tilde{f}_{-4}
                \label{eqn:2h_2g_amplitudes_initial_plus_minus}\,,
	\end{align}
\end{subequations}
where in the last line we used momentum conservation to get a manifestly symmetric
expression in $3\leftrightarrow 4$.
It will be convenient to define the variables
\begin{equation}
	\tilde{s}_{ij}\equiv2p_{i}\cdot p_{j}\qquad\textrm{for\ }i\ne j\,.
\end{equation}
These satisfy, 
\begin{align}\label{eqn:2h_2g_momentum_conservation}
	&\tilde{s}_{12}=2m^{2}+\tilde{s}_{34}\,, \quad  \tilde{s}_{13}=\tilde{s}_{24}  \,, \quad   \tilde{s}_{23}=\tilde{s}_{14}  \,, \quad \tilde{s}_{12} +\tilde{s}_{13} + \tilde{s}_{23} = 0\,.
\end{align}
The functions $f_{-2}$, $\tilde{f}_{-4}$ can then be written as power series
of the form,
\beq\label{eqn:2h_2g_generic_expansion_term_symmetric}
 a_{n,k,l}\,
 \tilde{s}_{12}^{n}\tilde{s}_{13}^{k}\tilde{s}_{23}^{l} \,
 =a_{n,k,l} \tilde{s}_{12}^{n}
 \left(\frac{\tilde{s}_{13}+\tilde{s}_{24}}{2}\right)^{k}
 \left(\frac{\tilde{s}_{23}+\tilde{s}_{14}}{2}\right)^{l}\,,
\eeq
where the coefficients $a_{n,k,l}$ satisfy, $a_{n,k,l} = a_{n,l,k}$.

\subsubsection{The $3h$ vertex}
In order to obtain the factorizable parts of the $2h + 2g$ amplitudes we will
need the three-point vertex for three massive scalars.
It is easy to show that this is a constant.
Let us label the three massive momenta by $p_{1}\,, p_{2} $ and $p_{3}$, where $p_{i} = m^{2}_{i}$ for $i=1,2,3$.
For simplicity we take all three masses to be identical, $m_{i} \equiv m$.
The amplitude does not carry any little group weights or indices and has mass dimension one. It is therefore a function of 
$\tilde{s}_{ij}$, with $i,j=1,2,3$. Since $\tilde{s}_{ij}=m^2$ the amplitude is a
constant, which we label $c_{3h}$.

\subsubsection{The $hh;++$ amplitude}
Let us first consider the factorizable part (see~\figref{2h_2g_factorization_dim_5_three_point_higgs}).
Gluing the scalar three-point vertex $c_{3h}$ and the $h\to 2g$ amplitude from~\eqref{h_2g_final_amp} we get
\begin{align}\label{eqn:2h_2g_plus_plus_factorization}
	\mathcal{M}\parn{2h; g^{a+}\parn{p_{1}} g^{b+}\parn{p_{2}}} & = \delta^{ab} \, \frac{i c_{3h}\,  c^{hgg}_{5}}{\Lambda}  \, \frac{\sqb{12}^{2}}{\tilde{s}_{12} - m^{2}}\,.
\end{align}
We can now turn to the non-factorizable parts.
Expanding $f_{-2}$ in terms of
$\left\{ \tilde{s}_{12}\,,\tilde{s}_{13}\,,\tilde{s}_{23}\right\} $
up to dimension 10, 
and maintaining symmetry under $1\leftrightarrow2$ and $3 \leftrightarrow 4$,
we get,
\begin{align}\label{eqn:2h_2g_plus_plus_up_to_dim_10}
	\mathcal{M}\parn{2h; g^{a+}\parn{p_{1}} g^{b+}\parn{p_{2}}} & = \delta^{ab} \, \frac{\sqb{12}^{2}}{\Lambda^{2}}  \, \bigg[
	a^{hh}_{0,0,0} + \frac{\left(\tilde{s}_{13}+\tilde{s}_{23}\right) a^{hh}_{0,0,1}}{\Lambda ^2}+\frac{\tilde{s}_{12} a^{hh}_{1,0,0}}{\Lambda ^2} \nn\\
	& + \frac{\tilde{s}_{12}^2 a^{hh}_{2,0,0}}{\Lambda ^4}+\frac{\left(\tilde{s}_{13}+\tilde{s}_{23}\right) \tilde{s}_{12} a^{hh}_{1,0,1}}{\Lambda ^4}+\frac{\left(\tilde{s}_{13}^2+\tilde{s}_{23}^2\right) a^{hh}_{0,0,2}}{\Lambda ^4}+\frac{\tilde{s}_{13} \tilde{s}_{23} a^{hh}_{0,1,1}}{\Lambda ^4}
	\bigg]\,.
\end{align}
Using~\eqref{2h_2g_momentum_conservation} and adding the factorizable part from~\eqref{2h_2g_plus_plus_factorization}
we finally have, 
\begin{align}\label{eqn:2h_2g_plus_plus_up_to_dim_10_final}
  \mathcal{M}\parn{2h; g^{a+}\parn{p_{1}} g^{b+}\parn{p_{2}}} &
  = \delta^{ab} \, \sqb{12}^{2}  \, \bigg[
\frac{c_{3h}\,  c^{hgg}_{5}}\Lambda\, \frac{i}{\tilde{s}_{12} - m^{2}} +
    \frac{a^{hh}_{6}}{\Lambda^2} +\frac{a^{hh}_{8}}{\Lambda ^4} \, \tilde{s}_{12}
    + \frac{a^{hh}_{8}}{\Lambda^{6}} \, \tilde{s}_{12}^2 + \frac{a^{\prime hh}_{8}}{\Lambda^{6}} \, \tilde{s}_{13} \tilde{s}_{14} 
	 	\bigg]\,.
\end{align}

\subsubsection{The $hh;+-$ amplitude}
The factorizable part of the $hh;g^+g^-$ amplitude (see~\figref{2h_2g_factorization_dim_5})
is,
\begin{align}
  \mathcal{M}_{hgg}\parn{2^{b+} P^{c+}} \, \frac{-i\delta^{cd}}{P^{2}} \,
  \mathcal{M}_{hgg}\parn{1^{a-}\parn{-P}^{d-}} 
  &=
  \mathcal{M}_{hgg}\parn{2^{b+} P^{c+}} \, \frac{-i\delta^{cd}}{\tilde{s}_{13}+m^{2}} \, \mathcal{M}_{hgg}\parn{1^{a-}\parn{-P}^{d-}}\nn \\
	&=i\left(\frac{c_{5}^{hgg}}{2\Lambda}\right)^{2}\,\braKet 1{p_{3}-p_{4}}2^{2}\,\frac{\delta^{ab}}{\tilde{s}_{13}+m^{2}}\,,
\end{align}
where $P=p_{1}+p_{3}$  and we have symmetrized the helicity part.
Adding the crossed channel with $3 \to 4 $ we get,
\begin{align}\label{eqn:2h_2g_plus_minus_factorizable_part}
	\mathcal{M}\parn{2h; g^{a-}\parn{p_{1}} g^{b+}\parn{p_{2}}} & = i \delta^{ab} \, \braKet{1}{p_{3} - p_{4}}{2}^{2} \, \left(\frac{c_{5}^{hgg}}{2\Lambda}\right)^{2} \, \bigg[\frac{1}{\tilde{s}_{13}+m^{2}} + \frac{1}{\tilde{s}_{14}+m^{2}}\bigg] \nn \\
	 & =i\left(\frac{c_{5}^{hgg}}{2\Lambda}\right)^{2}\,\braKet 1{p_{3}-p_{4}}2^{2}\delta^{ab}\,\left[\frac{2m^{2}-\tilde{s}_{12}}{\left(\tilde{s}_{13}+m^{2}\right)\left(\tilde{s}_{23}+m^{2}\right)}\right]\,.
\end{align}
Turning to the non-factorizable part, we expand $\tilde{f}_{-4}$ of~\eqref{2h_2g_amplitudes_initial_plus_minus} as
\begin{align}
\label{eqn:2h_2g_plus_minus_final}
\mathcal{M}\parn{2h; g^{a-}\parn{p_{1}} g^{b+}\parn{p_{2}}}
&
= \delta^{ab} \, 
\frac{\braKet{1}{p_{3}-p_{4}}{2}^{2}}{\Lambda^{4}} \, \sum_{\substack{n,k,l=0;\\0\le n+k+l}}^{\infty}\frac{\tilde{s}_{12}^{n}\tilde{s}_{13}^{k}\tilde{s}_{23}^{l}}{\Lambda^{2\left(n+k+l\right)}} \, b^{hh}_{n,k,l}\,,
\end{align}
where $b^{hh}_{n,k,l} = b^{hh}_{n,l,k}$.
Computing the series in~\eqref{2h_2g_plus_minus_final} up to dimension 10
and adding the factorizable part 
from~\eqref{2h_2g_plus_minus_factorizable_part} we get
%
\beq
\label{eqn:2h_2g_plus_minus_up_to_dim_10_final}
\mathcal{M}\parn{2h; g^{a-}\parn{p_{1}} g^{b+}\parn{p_{2}}}
= \delta^{ab} \, \frac{\braKet{1}{p_{3}-p_{4}}{2}^{2}}{\Lambda^{4}} \, \bigg[
  b^{hh}_{8} + \frac{b^{hh}_{10}}{\Lambda^{2}} \, \tilde{s}_{12}
  -  \parn{\frac{c^{hgg}_{5} \Lambda}{2}}^{2} \, \left[\frac{2m^{2}-\tilde{s}_{12}}{\left(\tilde{s}_{13}+m^{2}\right)\left(\tilde{s}_{23}+m^{2}\right)}\right]
	\bigg]\,.
\eeq
        %

\subsubsection{Inferring the EFT}
\begin{table}[t]
	\centering
	\caption{Operators contributing to the $hhgg$ amplitudes.
          The number of independent operators is shown in brackets. \label{tab:EFT_2h_2g_operator_list_1}
         }
	\begin{tabular}{|c|c|c|}
		\hline 
		\multirow{2}{*}{Mass dimension} & \multicolumn{2}{c|}{Operators}\\
		\cline{2-3} 
		& $\mathcal{M}\parn{++}$ & $\mathcal{M}\parn{+-}$\\
		\hline 
		\hline 
		6 & $\GSD^{2} \, h^{2}$~[1] & --- \\
		\hline 
		8 & $\mathcal{D}^{2} \, \GSD^{2} \, h^{2}$~[1] &
                $\mathcal{D}^{2} \, \GSD \, \GASD \, h^{2}$~[1]\\
		\hline 
		10 & $\mathcal{D}^{4} \, \GSD^{2} \, h^{2}$~[2]
                & $\mathcal{D}^{4} \, \GSD \, \GASD \, h^{2}$~[1] \\
		\hline 
	\end{tabular}
\end{table}
The EFT contributions to the $hh;++$ and $hh;--$ amplitudes are given
in~\eqsref{2h_2g_plus_plus_up_to_dim_10_final}{2h_2g_plus_minus_up_to_dim_10_final}.
Aside from the constants $c^{hgg}_{5}$, which controls the $h\to 2g$ amplitudes, and $c_{3h}$, which controls the $3h$ vertex,
the $hh;++$ amplitude depends on four dimensionless constants,
while the $hh;+-$ amplitude depends on two dimensionless constants.
Therefore the $hh;++$ amplitude gets contributions from one operator 
at dimension 6, one operator at dimension 8,
and two operators at dimension 10.
The $hh;+-$ amplitude gets contributions from
one operator at dimension 8 and one at dimension 10. 
This agrees with the counting obtained by using the Mathematica notebook
of~\cite{Henning:2015alf}.
The results are summarized in Table~\ref{tab:EFT_2h_2g_operator_list_1}.

\section{The vector plus three gluon amplitudes~${\cal M}(Z^\prime;ggg)$}
\label{sec:spin1}
In this section we consider amplitudes with a single massive spin-1
particle, which we denote by $Z^\prime$,  and three gluons. 
We will then use these to obtain the massless vector amplitudes.
Yang's theorem forbids the massive vector decay to two gluons.
For completeness, we review the derivation of Yang's theorem via
on-shell methods following~\cite{Arkani-Hamed:2017jhn}.
This will also serve as a simple illustration of the 
spinor helicity formalism for massive particles of~\cite{Arkani-Hamed:2017jhn}.

An amplitude with an external massive spin-1 particle can be decomposed as~\cite{Arkani-Hamed:2017jhn} 
\begin{align}\label{eqn:massive_spin_1_general_amplitude_decomposition}
	\mathcal{M}^{I_{1}I_{2}}&= \bs{\lambda}_{\alpha_{1}}^{I_{1}}\bs{\lambda}_{\alpha_{2}}^{I_{2}}M^{\left\{ \alpha_{1},\alpha_{2}\right\} }=\tilde{\bs{\lambda}}_{\dot{\alpha}_{1}}^{I_{1}}\tilde{\bs{\lambda}}_{\dot{\alpha}_{2}}^{I_{2}}\tilde{M}^{\left\{ \dot{\alpha}_{1},\dot{\alpha}_{2}\right\} } \,,
\end{align}
where the boldface spinors $\bs\lambda$ are spinor helicity variables
for the massive
momentum, (see~\appref{notation}),
and $I_{1},I_{2}=1,2$ are SU(2) indices of the little group
of the massive particle.
The reduced amplitudes, $M$ and $\tilde{M}$, are  symmetric in the
$\SL{2}{\mathbb{C}}$ indices $\alpha$ and $\dot\alpha$ respectively.
The dotted and undotted representations of the amplitude are equivalent;
we may use the massive momentum 
to switch between the dotted and undotted spinors (see~\eqref{spinor_helicity_variables_massive_momentum_rels_between_half_brackets}).

The $Z^\prime$ plus two gluon reduced amplitude can then be written as,
\begin{align}\label{eqn:massive_spin_1_2g_ansatz_dotted}
  \tilde{M}^{\dot{\alpha}_{1}\dot{\alpha}_{2}}
  \Big(Z^\prime(\bs{p});g(p_1) g(p_2)\Big)
	&=\left[12\right]^{n_{12}}
	\left(\tilde{\lambda}_1^{n_1} \tilde{\lambda}_2^{n_2}\right)^{\dot{\alpha}_{1}\dot{\alpha}_{2}} 
	 f_{-\ell}\left(s_{12},\Lambda\right)
	\,,
\end{align}
where $\tilde\lambda_{i=1,2}$ is the spinor associated with the
gluon momentum $p_i$, and $n_i=0,1,2$ such that $n_1+n_2=2$.
Requiring the correct little group weights and dimensionality gives
the relations
\beq
\label{eqn:massive_spin1_2g_helicity_mass_eqs}
	n_{1}+n_{12} = 2h_{1}\,,~~
n_{2}+n_{12} = 2h_{2}\,,~~
n_{12} -\ell +2 = 1\,,
\eeq
so that the full amplitude is  given by 
\begin{align}\label{eqn:massive_spin1_2g_final_amp}
  \mathcal{M}\Big(Z^\prime(\bs{p});g^{a, h_1}(p_1) g^{b, h_2}(p_2)\Big)
  = \delta^{ab}\,\sqb{12}^{h-1}\sqb{1\bs{p}}^{2h_{1} + 1-h}\sqb{2\bs{p}}^{2h_{2} + 1-h}f_{-h}\parn{m^2, \Lambda}\,,
\end{align}
where $h=h_1+h_2$,  and $a, b$ are color indices.
We have also employed the ``\textbf{bold}'' notation of~\cite{Arkani-Hamed:2017jhn}, where the SU(2) index of the $\bs{p}$-spinor is suppressed,
with the understanding that the amplitude is  symmetric in these indices. 
We see that the amplitude~\eqref{massive_spin1_2g_final_amp}
is   antisymmetric under $1 \leftrightarrow 2$ exchange.
We conclude that it must vanish, in accord with the spin-statistics theorem.

The amplitude $Z^{\prime}_{\mu} \to 3g$ is therefore the first non-zero amplitude,
and has no factorizable part. 
We choose the following ansatz for the  reduced amplitude,
\begin{align}\label{eqn:spin1_ansatz_amplitudes_initial_dotted}
  &\tilde{M}^{\dot{\alpha}_{1}\dot{\alpha}_{2}}
   \Big(Z^\prime(\bs{p_4});g(p_1)^{h_1} g(p_2)^{h_2} g(p_2)^{h_3}  \Big)
= \sqb{12}^{n_{12}} \sqb{13}^{n_{13}} \sqb{23}^{n_{23}}  \left(\tilde{\lambda}_{1}^{n_{1}}\tilde{\lambda}_{2}^{n_{2}}\tilde{\lambda}_{3}^{n_{3}}\right)^{\dot{\alpha}_{1}\dot{\alpha}_{2}} \, f_{-\ell}\parn{s_{12}, s_{13}, s_{23}} \,, 
\end{align}
where $n_{i} = 0,1,2$ with $\sum_{i=1}^{3}n_{i} = 2$ and for now we suppressed the color indices. 
Requiring the correct helicity weights and mass dimension in~\eqref{spin1_ansatz_amplitudes_initial_dotted} gives,
\begin{align}\label{eqn:spin1_mass_eqn_dotted_amp}
	\sum^{3}_{1=i<j} n_{ij} = \ell - 2\,,
\end{align}
and
\begin{align}\label{eqn:spin1_helicity_eqns_dotted_amp}
		n_{1} + n_{12} + n_{13} = 2h_{1} \,,~~
		n_{2} + n_{12} + n_{23} = 2h_{2} \,,~~
		n_{3} + n_{13} + n_{23} = 2h_{3} \,.
\end{align}
\eqsref{spin1_mass_eqn_dotted_amp}{spin1_helicity_eqns_dotted_amp}
give the selection rule,
\begin{align}\label{eqn:spin1_mass_eqn_sol_dotted}
	\ell &=1+h_{1}+h_{2}+h_{3}\,,
\end{align}
with
\begin{align}\label{eqn:spin1_mass_helicity_sol_dotted}
	&n_{12}=n_{3}+h_{1}+h_{2}-h_{3}-1 \eqCS n_{23}=-n_{2}-n_{3}-h_{1}+h_{2}+h_{3}+1 \,, \nn \\ 
	&n_{13}=n_{2}+h_{1}-h_{2}+h_{3}-1 \eqCS n_{1}=-n_{2}-n_{3}+2\,.
\end{align}
%

\subsection{The $Z^\prime;+++$ amplitude}
Substituting $h_1 = h_2 = h_3 = +1$ in~\eqsref{spin1_mass_eqn_sol_dotted}{spin1_mass_helicity_sol_dotted} we obtain 
\begin{align}
\label{eqn:massive_spin_1_ppp_initial_res}
	&\mathcal{M}\parn{Z^{\prime}(\bs{p_4});g^{a+}\parn{p_1};g^{b+}\parn{p_2};g^{c+}\parn{p_3}} = C^{abc}\bigg[ 
	\sqb{23}^{2}\sqb{1\bs{4}}^{2} f^{\prime\pm}_{-4}\parn{2;3} 
	+ \sqb{12}^{2}\sqb{3\bs{4}}^{2} f^{\prime\pm}_{-4}\parn{1;2}
      \nn  \\
	     &+ \sqb{13}^{2}\sqb{2\bs{4}}^{2} f^{\prime\pm}_{-4}\parn{1;3} 
	      + \sqb{12}\sqb{13}\sqb{2\bs{4}}\sqb{3\bs{4}}
             f^{\pm}_{-4}\parn{2;3} 
	  + \sqb{13}\sqb{23}\sqb{1\bs{4}}\sqb{2\bs{4}} f^{\pm}_{-4}\parn{1;2} \nn  \\
     &+ \sqb{12}\sqb{23}\sqb{1\bs{4}}\sqb{3\bs{4}} f^{\pm}_{-4}\parn{1;3} \bigg] 
\end{align}
where $C^{abc}=f^{abc}$ or $d^{abc}$ and $f_{-\ell}\parn{i;j},\ f_{-\ell}^{\prime}\parn{i;j}$ are analytic
functions of the $s_{ij}$'s of mass dimension $-\ell$.
For $C^{abc}=d^{abc}$ ($C^{abc}=f^{abc}$) these functions are symmetric (antisymmetric)
under $i\leftrightarrow j$, and we use the superscript $\pm$ to denote
the symmetric and antisymmetric versions respectively. 
Using Schouten identities we can eliminate the first three terms to
obtain\footnote{The Schouten identity reads
	$\sqb{12}\sqb{34} + \sqb{13}\sqb{42} + \sqb{14}\sqb{23} = 0$,
	and similarly for the angle brackets.},
\begin{align}\label{eqn:massive_spin_1_ppp}
  \mathcal{M} & \parn{Z^{\prime}(\bs{p_4});
    g^{a+}\parn{p_1};g^{b+}\parn{p_2};g^{c+}\parn{p_3}}=\\
  & d^{abc}\bigg[ \sqb{12}\sqb{23}\sqb{1\bs{4}}\sqb{3\bs{4}} f^+_{-4}\parn{1;3}  
	 -\sqb{13}\sqb{23}\sqb{1\bs{4}}\sqb{2\bs{4}} f^+_{-4}\parn{1;2} 
	 -\sqb{12}\sqb{13}\sqb{2\bs{4}}\sqb{3\bs{4}} f^+_{-4}\parn{2;3}
	 \bigg]\nn\\
  +&f^{abc}\bigg[ \sqb{12}\sqb{23}\sqb{1\bs{4}}\sqb{3\bs{4}} f^-_{-4}\parn{1;3}  
	 +\sqb{13}\sqb{23}\sqb{1\bs{4}}\sqb{2\bs{4}} f^-_{-4}\parn{1;2} 
	 +\sqb{12}\sqb{13}\sqb{2\bs{4}}\sqb{3\bs{4}} f^-_{-4}\parn{2;3}
	 \bigg]\,,\nn 
\end{align}
with
\begin{align}\label{eqn:massive_spin1_ppp_f}
  f^+_{-4}\parn{1;2}
  	&= \frac{1}{\Lambda^{4}} \, \parn{ c^{\parn{1}}_{8} + \frac{s_{12}}{\Lambda^{2}} \, c^{\parn{1}}_{10} +\frac{1}{\Lambda^{4}}\parn{s^{2}_{12} \, c^{\parn{1}}_{12} + s_{13}s_{23} \, c^{\parn{2}}_{12} }}   + \mathcal{O}\parn{\Lambda^{-10}}\,,\\
 	f^-_{-4}\parn{1;2} 
	&= \frac{s_{13} - s_{23}}{\Lambda^{6}} \parn{ c^{\parn{2}}_{10} + \frac{c^{\parn{3}}_{12}}{\Lambda^{2}} \,  s_{12}} + \mathcal{O}\parn{\Lambda^{-10}}\,,
\end{align}
with the remaining functions obtained by permutations of $\{1,2,3\}$.
Thus the amplitude involves six independent coefficients.
One coefficient with $d^{abc}$ at dimension 8,
two coefficients at dimension 10---one for each color structure,
and three coefficients at dimension 12---two with $d^{abc}$ and one with $f^{abc}$.

We can now infer the structure and number of the EFT operators which contribute to this amplitude. Using the EOM for $Z^{\prime}$,
we can write these operators in terms of field strengths only.
The different operators then contain three powers of $G_{\mu\nu}$ and a single power of 
$Z^\prime_{\mu\nu}$, with some number of derivatives.
There is one operator at dimension 8, which is schematically
$Z^\prime \SD{G}^3$, with $Z^\prime$ denoting the $Z^\prime$ field strength,
and with color indices contracted with $d^{abc}$,
two operators at dimension 10, namely $D^2 Z^\prime \SD{G}^3$,
one with $f^{abc}$ and
one with $d^{abc}$, and three operators of the form $D^4 Z^\prime \SD{G}^3$ at dimension
12---two with $d^{abc}$ and one with $f^{abc}$.
The effects of the lowest-order operators were considered
in~\cite{Bramante:2011qc,Alwall:2012np,Kumar:2012ba}.

\eqref{massive_spin_1_ppp} describes the
three $Z^\prime$ polarizations.
The positive polarization is obtained by taking
$\Ket{4^{I=1}}\Ket{4^{J=1}}$, the negative polarization is obtained by taking
$I=J=2$,
and the longitudinal polarization corresponds to the symmetric combination
of $I=1,J=2$.

Above we have chosen to write the amplitude as a sum of permutations 
over the spinor factors.
Alternatively, we can write it such that the functions of the $s_{ij}$'s
are manifestly symmetric, with
\begin{align}\label{eqn:massive_spin_1_ppp_n}
  &\mathcal{M}\parn{Z^{\prime}(\bs{p_4});g^{a+}\parn{p_1};g^{b+}\parn{p_2};g^{c+}\parn{p_3}}
  =  \nn\\
  &\sqb{12}\sqb{23}\sqb{31}\times \sqb{\bs{4}1}\anb{12}\sqb{2\bs{4}}\times
  \Bigg[f^{abc}\,\left(
  \frac1{s_{12}} f_{-4}^-(1;2)-\frac1{s_{13}} f_{-4}^-(1;3)-
  \frac1{s_{23}} f_{-4}^-(3;2)
  \right)\nn\\  
  & -d^{abc}\,\left(
  \frac1{s_{12}} f_{-4}^+(1;2)+ \frac1{s_{13}} f_{-4}^+(1;3)+
  \frac1{s_{23}} f_{-4}^+(3;2)
  \right)\Bigg] \nn\\
&+ m \, d^{abc}\, \sqb{12}\sqb{23}\sqb{31} \, \parnS{\frac{\left\langle 2\bs 4\right\rangle \sqb{2\bs 4}}{s_{23}}f_{-4}^{+}\parn{2;3}-\frac{\sqb{1\bs 4}\left\langle 1\bs 4\right\rangle }{s_{13}}f_{-4}^{+}\parn{1;3}}  \nn\\
&- m \, f^{abc}\, \sqb{12}\sqb{23}\sqb{31} \, \parnS{\frac{\left\langle 2\bs 4\right\rangle \sqb{2\bs 4}}{s_{23}}f_{-4}^{-}\parn{2;3}+\frac{\sqb{1\bs 4}\left\langle 1\bs 4\right\rangle }{s_{13}}f_{-4}^{-}\parn{1;3}}\,.
\end{align}
Using momentum conservation it is easy to check that the spinor prefactor
$\sqb{12}\sqb{23}\sqb{31}\times \sqb{\bs{4}1}\anb{12}\sqb{2\bs{4}}$
is actually completely symmetric under exchange of 1,2,3, up to terms of
order $m$. This form of the amplitude is convenient for some purposes as
we will see below.

Finally, recall that we chose to work with the dotted-indices reduced
amplitude, \eqref{spin1_ansatz_amplitudes_initial_dotted}.
Using the undotted reduced amplitude would give an equivalent solution.
We will return to this point below when we discuss the massless limit.

\subsection{The $Z^\prime;--+$ amplitude}
We now turn to the choice $h_1 = h_2 = -h_3 = -1$ for the gluon helicities.
Using~\eqref{spin1_mass_helicity_sol_dotted} the amplitude is, 
\begin{align}
  &\mathcal{M}\parn{Z^{\prime}(\bs{p_4});g^{a-}\parn{p_1};g^{b-}\parn{p_2};g^{c+}\parn{p_3}} =  d^{abc}
  \bigg\{
    \frac{\sqb{13}\sqb{23}\sqb{1\bs{4}}\sqb{2\bs{4}}}{\sqb{12}^{4}}A_{0}^{+}\parn{1;2}+\frac{\sqb{3\bs{4}}^{2}}{\sqb{12}^{2}}D_{0}^{+}\parn{1;2}\nn\\
& +\frac{1}{\sqb{12}^{4}}\bigg[\parn{\sqb{23}^{2}\sqb{1\bs{4}}^{2}+\sqb{13}^{2}\sqb{2\bs{4}}^{2}}B_{0}^{+}\parn{1;2}+\parn{\sqb{23}^{2}\sqb{1\bs{4}}^{2}-\sqb{13}^{2}\sqb{2\bs{4}}^{2}}B_{0}^{-}\parn{1;2}\bigg]\nn\\
    & +\frac{\sqb{3\bs{4}}}{\sqb{12}^{3}}\bigg[\parn{\sqb{23}\sqb{1\bs{4}}+\sqb{13}\sqb{2\bs{4}}}C_{0}^{-}\parn{1;2}+\parn{\sqb{23}\sqb{1\bs{4}}-\sqb{13}\sqb{2\bs{4}}}C_{0}^{+}\parn{1;2}\bigg] \bigg\}\,,
\end{align}
where
$A^{+}_{0}\parn{1;2}\,, B^{\pm}_{0}\parn{1;2} \,, C^{\pm}_{0}\parn{1;2}\,, D^{+}_{0}\parn{1;2}$
are all dimensionless, analytic functions of the $s_{ij}$'s,
which are symmetric or anti-symmetric under $1\leftrightarrow2$.
The symmetric (antisymmetric) functions are denoted by a $+ \ \parn{-}$ superscript. 
Using Schouten identities we can rewrite the full amplitude as,
\begin{align}\label{eqn:massive_spin_1_mmp_final}
&\mathcal{M}\parn{Z^{\prime};g^{a-}\parn{p_1};g^{b-}\parn{p_2};g^{c+}\parn{p_3}}  
=d^{abc}\bigg[
\anb{12}^{2}\sqb{3\bs{4}}^{2}D^{+}_{-4}\parn{1;2}\nn\\
&+\anb{12}^{4}\sqb{13}\sqb{23}\sqb{1\bs{4}}\sqb{2\bs{4}}A^{+}_{-8}\parn{1;2}
+\anb{12}^{3}\sqb{3\bs{4}}\Big(
  \sqb{23}\sqb{1\bs{4}}
  +\sqb{13}\sqb{2\bs{4}}\Big)\,
C^{-}_{-6}\parn{1;2}\bigg] \nn\\
&f^{abc}\bigg[
\anb{12}^{2}\sqb{3\bs{4}}^{2}D^{-}_{-4}\parn{1;2}
+\anb{12}^{4}\sqb{13}\sqb{23}\sqb{1\bs{4}}\sqb{2\bs{4}}A^{-}_{-8}\parn{1;2}
\nn\\&+\anb{12}^{3}\sqb{3\bs{4}}
\Big(
  \sqb{23}\sqb{1\bs{4}}
  +\sqb{13}\sqb{2\bs{4}}\Big)\,
C^{+}_{-6}\parn{1;2}\bigg]\,,
\end{align}
with the coefficient functions,
\begin{align}\label{eqn:mmpfn_in}
	D_{-4}^{+}&=\frac{d_{8}}{\Lambda^{4}}+\frac{d_{10}^{\parn 1}}{\Lambda^{6}}
	\,s_{12}+\frac{d_{12}^{\parn 1}s_{12}^{2}
	  +d_{12}^{\parn 2}s_{13}s_{23}}{\Lambda^{8}}\,,~~~~
 	D_{-4}^{-}=\parn{s_{23}-s_{13}}\parn{\frac{d_{10}^{\parn 2}}{\Lambda^{6}}
          +\frac{d_{12}^{\parn 5}s_{12}}{\Lambda^{8}}}\,,\nn\\       
	A_{-8}^{+}&=\frac{d_{12}^{\parn 3}}{\Lambda^{8}}\,,~~~~
        A_{-8}^{-}=0 \,,\\
   	C_{-6}^{+}&=\frac{d_{10}^{\parn 3}}{\Lambda^{6}}
        +\frac{d_{12}^{\parn 6}}{\Lambda^{8}}\,s_{12}\,,~~~~     
	C_{-6}^{-}=\parn{s_{23}-s_{13}}
	\frac{d_{12}^{\parn 4}}{\Lambda^{8}} \,. \nn
\end{align}
Again, this form of the amplitude encapsulates all three $Z^\prime$ polarizations.
The number of independent operators can  be read
from~\eqref{mmpfn_in}. 
There is one operator at dimension 8, of the form $Z^\prime G^3$, with color indices contracted with $d^{abc}$, 
and three operators of the form $D^2 Z^\prime G^3$ at dimension 10;
two with $d^{abc}$  and one with $f^{abc}$.
At dimension 12 there are
six operators of the form $D^4 Z^\prime G^3$;
five with $d^{abc}$ and one with $f^{abc}$.

\subsection{The massless $Z^\prime$ amplitudes}
Amplitudes with a massless $Z^\prime$ can be calculated directly, much like we did for the massive $Z^\prime$.
Recall that our starting point above was the ``dotted''
amplitude~\eqref{spin1_ansatz_amplitudes_initial_dotted}.
In the massive case, the dotted and undotted solutions are equivalent,
since the different $Z^\prime$ polarizations are related.
For a massless $Z^\prime$,
the dotted form we used above gives the positive-helicity $Z^\prime$,
while the undotted form gives the negative-helicity  $Z^\prime$.
Other than that, the derivation would proceed just as above, with the 4-spinors
unbolded, and with $m=0$.

It is instructive however to obtain the massless amplitudes by taking the massless limit of the amplitudes we already calculated.
Recall that~\eqsref{massive_spin_1_ppp}{massive_spin_1_mmp_final} describe the
three $Z^\prime$ polarizations.
Thus the massive-$Z^\prime$ amplitudes  split into the positive and negative
massless-$Z^\prime$ helicities, plus the longitudinal polarization,
which should coincide with
the scalar plus 3-gluon amplitudes, ${\cal M}(h;ggg)$.
In the massless limit, $\Ket{4^{I=2}}$ goes to zero (see~\appref{helimit}),
so only the $Z^{\prime+}$ amplitudes survive
in~\eqsref{massive_spin_1_ppp}{massive_spin_1_mmp_final}.
This is not surprising, since we used the dotted-indices ansatz~\eqref{spin1_ansatz_amplitudes_initial_dotted}, which
reduce to the positive-helicity $Z^\prime$ amplitudes
in the massless limit.
It is therefore useful to rewrite the amplitudes such that both $\Ket{4^{I=1}}$
and $\ket{4^{I=2}}$ appear, and all three polarizations are transparent
in the massless limit.
Terms with $\Ket{4}\Ket{4}$ then describe the positive polarization as before, while
the negative polarization corresponds to terms with $\ket{4}\ket{4}$,
and the longitudinal polarization corresponds to mixed terms with $\Ket{4}\ket{4}$.
The different polarizations should appear with different suppressions of $\Lambda$,
since in the massless limit the net helicity of the operator is related to its dimension.
Another way to say this is that amplitudes with different $Z^\prime$
helicities are related by different factors
of the mass $m$, corresponding to helicity flips on the $Z^\prime$ leg.

To rewrite the massive amplitudes~\eqsref{massive_spin_1_ppp}{massive_spin_1_mmp_final}, we use the momentum $p_{4}$ to
convert between $\Ket{4}$ and $\ket{4}$
(see~\eqref{spinor_helicity_variables_massive_momentum_rels_between_half_brackets} and~\appref{helimit}).
Consider first ${\cal M}(Z^\prime;--+)$.
Simply taking the massless limit in~\eqref{massive_spin_1_mmp_final},
the three different spinor structures  collapse into a single structure,
corresponding to the positive $Z^\prime$ helicity.
Using~\eqref{massive_spinor_rels} we then rewrite the amplitude as,
\begin{align}\label{eqn:massive_spin_1_mmp_HelicityExplicit_final1}
&\mathcal{M}\parn{Z^{\prime};g^{a-}\parn{p_1};g^{b-}\parn{p_2};g^{c+}\parn{p_3}}  
\\
&=d^{abc}\, \anb{12}^{2}\times 
\bigg[
\sqb{3\bs{4}}^{2} \tilde{f}_{-4}^+\parn{1;2}  
+\sqb{13}\sqb{23}\anb{1\bs{4}}\anb{2\bs{4}}\tilde{f}_{-6}^+\parn{1;2}
+\sqb{3\bs{4}}\parn{\sqb{31}\anb{1\bs{4}}
  - \sqb{32}\anb{2\bs{4}}} \tilde{f}_{-5}^{-}\parn{1;2} 
\bigg]\nn\\
&+f^{abc}\, \anb{12}^{2}\times 
\bigg[
\sqb{3\bs{4}}^{2} \tilde{f}_{-4}^-\parn{1;2}  
+\sqb{13}\sqb{23}\anb{1\bs{4}}\anb{2\bs{4}}\tilde{f}_{-6}^-\parn{1;2}
+\sqb{3\bs{4}}\parn{\sqb{31}\anb{1\bs{4}}
  - \sqb{32}\anb{2\bs{4}}} \tilde{f}_{-5}^{+}\parn{1;2} 
\bigg]\nn\,,
\end{align}
where we redefined the coefficient functions with,
\begin{align}\label{eqn:mmpfn}
	\tilde{f}_{-4}^{+}(1;2)&=\frac{d_{8}}{\Lambda^{4}}+\frac{d_{10}^{\parn 1}}{\Lambda^{6}}
	\,s_{12}+\frac{d_{12}^{\parn 1}s_{12}^{2}
		+d_{12}^{\parn 2}s_{13}s_{23}}{\Lambda^{8}}
	\,,~~~~
	\tilde{f}_{-4}^{-}(1;2)=\parn{s_{23}-s_{13}}
	\Big(\frac{d_{10}^{(3)}}{\Lambda^{6}} +
	\frac{d_{12}^{(4)}}{\Lambda^{8}} s_{12}
	\Big)\,,
	\nn\\
	\tilde{f}_{-5}^{+}(1;2)&=  \frac{m \, d_{10}^{\parn 2}}{\Lambda^{6}}+
	\frac{m \, d_{12}^{\parn 3}}{\Lambda^{8}}  s_{12}
	\,,~~~~
	\tilde{f}_{-5}^{-}(1;2)=\parn{s_{13}-s_{23}}\frac{m \, d_{12}^{\parn 5}}{\Lambda^{8}} \,, \nn\\
	\tilde{f}^{+}_{-6}\parn{1;2} &=
        \frac{m^{2} \, s_{12} \, d^{\parn{6}}_{12}}{\Lambda^{8}} \eqCS 
	\tilde{f}^{-}_{-6}\parn{1;2} = 0 \,, 	
\end{align}
where $\tilde{f}^{\pm}_{-6} = m^{2} A^{\pm}_{-8}$
and $\tilde{f}^{\pm}_{-5} = m C^{\pm}_{-6}$. 
It is now easy to see the three spin polarizations both in the massive and
in the massless case.
The positive helicity $Z^\prime$ is given by terms with two $\Ket{4}$ spinors, the negative helicity is given by terms with two $\ket{4}$ spinors,
and the longitudinal polarization corresponds to the mixed terms.

In the massless limit, only $\Ket{4^1}$ and $\ket{4^2}$ survive, so we can
simply unbold the 4-spinors
in~\eqref{massive_spin_1_mmp_HelicityExplicit_final1}.
Furthermore, explicit factors of $m$ which appear in the coefficient functions
of~\eqref{mmpfn} can be absorbed by rescaling $\Lambda$.
Thus for example, we can rewrite
$m^{2} \, s_{12} \, d^{(6)}_{12}/\Lambda^{8}$ as
$s_{12} \, d^{(6)}_{12}/\tilde\Lambda^6$.
This contribution is then associated with a dimension-10 operator.
This is just as expected. The mass $m$ is not a parameter of the massless
theory, 
and $Z^{\prime+}$ and $Z^{\prime-}$ amplitudes are generated by operators of
different dimensions.
Examining~\eqref{massive_spin_1_mmp_HelicityExplicit_final1}, we see that
the $\sqb{34}^2$ terms give  
$\mathcal{M}(Z^{\prime +};--+)$, while the $\anb{14}\anb{24}$ terms give 
 $\mathcal{M}(Z^{\prime -};--+)$.
Thus the massless
$\mathcal{M}(Z^{\prime +};--+)$
amplitude gets contributions from $\tilde{f}_{-4}^\pm$.
There is one operator with $d^{abc}$ at dimension 8, two operators (one with $d^{abc}$ and one with $f^{abc}$ at dimension 10, and three operators at dimension 12 (two with $d^{abc}$ and one with $f^{abc}$).
The $\mathcal{M}(Z^{\prime -};--+)$
amplitude gets contributions from $\tilde{f}_{-6}^\pm$.
There is one operator with $d^{abc}$ at dimension 10 and two operators,
one with $d^{abc}$ and one with $f^{abc}$ at dimension 12.
In this case too we used the Mathematica notebook of~\cite{Henning:2015alf} to derive the EFT operators with a massless
vector plus three gluon operators, and checked that the results agree.
We summarize the EFT operators and their numbers in~\tableref{EFT_X_3g_operator_list}.
%
\begin{table}[t]
	\centering
	\caption{Operators contributing to the $Z^\prime ggg$ amplitude for
          a massless $Z^\prime$.
          The numbers of independent operators per each color
          structure are shown in brackets.
          (In the Table, we use $Z^\prime$ to denote the field strength).
		\label{tab:EFT_X_3g_operator_list}}	
		\begin{tabular}{|c|c|c|}
			\hline 
			\multirow{2}{*}{Mass dimension} & \multicolumn{2}{c|}{Operators}\\
			\cline{2-3} 
			& $\mathcal{M}\parn{+++} $ & $\mathcal{M}\parn{--+}$\\
			\hline 
			\hline 
			8 & $ \GSD^{3} \, \SD{{Z^\prime}} \,  \parnS{1, d^{abc}} $ &  $\GASD^{2} \, \GSD \, \SD{{Z^\prime}} \, \parnS{1, d^{abc}}$\\
			\hline 
			10 & \makecell{ $\mathcal{D}^{2} \, \GSD^{3} \, \SD{{Z^\prime}} \, \parnS{1, d^{abc}; 1, f^{abc}} $ \\ $ \mathcal{D}^{2} \, \GSD^{3} \, \ASD{{Z^\prime}} \, \parnS{1, d^{abc}}$ }
			&  \makecell{ $\mathcal{D}^{2} \, \GASD^{2} \, \GSD \, \ASD{{Z^\prime}} \, \parnS{1, d^{abc}} $\\$ \mathcal{D}^{2} \, \GASD^{2} \, \GSD \, \SD{{Z^\prime}} \, \parnS{1, d^{abc}; 1, f^{abc}} $} \\
			\hline 
			12 & $\mathcal{D}^{4} \, \GSD^{3} \, \SD{{Z^\prime}} \, \parnS{1, d^{abc}}$ 
			& \makecell{$\mathcal{D}^{4} \, \GASD^{2} \, \GSD \, \ASD{{Z^\prime}} \, \parnS{1, d^{abc}; 1, f^{abc}} $\\$ \mathcal{D}^{4} \, \GASD^{2} \, \GSD \, \SD{{Z^\prime}} \, \parnS{2, d^{abc}; 1, f^{abc}}$} \\
			\hline 
	\end{tabular}
\end{table}

Finally, we can verify the relation between the amplitude
with a longitudinally-polarized $Z^\prime$
and  the scalar amplitude ${\cal M}(h;g^-g^-g^+)$.
The former is easily read-off~\eqref{massive_spin_1_mmp_HelicityExplicit_final1}.
Only the $f^{abc}$ term contributes at this order,
\begin{align}
&\mathcal{M}\parn{Z^{\prime 0};g^{a-}\parn{p_1};g^{b-}\parn{p_2};g^{c+}\parn{p_3}}  
  = \anb{12}^{2}\sqb{34}\Big[
    \sqb{31}\anb{14} - \sqb{32}\anb{24}\Big]
  \parn{\tilde{f}^+_{-5}\parn{1;2}
    \, f^{abc} + \tilde{f}^-_{-5}\parn{1;2} \, d^{abc}}
\nn\\
&= \anb{12}^{2}\sqb{31}\sqb{32}\anb{12} \parn{\tilde{f}^+_{-5}\parn{1;2} \, f^{abc} + \tilde{f}^-_{-5}\parn{1;2} \, d^{abc}} \nn\\
&=  \frac{\anb{12}^3}{\anb{13}\anb{23}} s_{13} s_{23} \parn{\tilde{f}^+_{-5}\parn{1;2} \, f^{abc} + \tilde{f}^-_{-5}\parn{1;2} \, d^{abc}} \nn \\
&= \frac{\anb{12}^3}{\anb{13}\anb{23}}s_{13} s_{23}\left(f^{abc} \parn{\frac{m \, d_{10}^{\parn 2}}{\Lambda^{6}}+
	\frac{m \, d_{12}^{\parn 3}}{\Lambda^{8}}  s_{12}} + \frac{d^{abc} \, d_{12}^{\parn 5} \, m \,\parn{s_{13} - s_{23}}}{\Lambda^{8}} + \ldots\right)\,,  
\end{align}
which agrees with the result~\eqref{h_3g_++-_final_result} upon replacing
angle brackets by square brackets, owing to the opposite gluon helicities
we have.
Furthermore, this contribution starts at $m/\Lambda^6$, namely at dimension-9,
which is where ${\cal M}(h;--+)$ first appears.
What we are seeing is nothing but the Higgs mechanism.
The longitudinal $Z^\prime$ polarization comes from the Goldstone boson,
so in the EFT operators generating the amplitudes we can replace $Z^\prime_\mu$ by
$\partial_\mu h$. Thus the relevant $h;--+$ amplitude must come from an operator
containing at least one derivative. The first such operator is
indeed the dimension-9 operator $D^2 h G^3$, which appears with $f^{abc}$
(see~\tableref{EFT_h_3g_operator_list}).

Let us now turn to the massless limit of the $Z^\prime; +++$ amplitude. 
In this case we find,
\begin{align}\label{eqn:zppp_zerom1}
  &\mathcal{M}\parn{Z^{\prime}; g^{a+}\parn{p_1};g^{b+}\parn{p_2};g^{c+}\parn{p_3}}
  =\\
&+
d^{abc}\,\parn{ c^{\parn{1}}_{8} \frac1{\Lambda^4}+
		c^{\parn{1}}_{10}\,\frac{s_{12}}{\Lambda^{6}} 
		+ c^{\parn{2}}_{12}\,\frac{s_{13}s_{23}}{\Lambda^8}
}
\sqb{13}\sqb{23}\sqb{1\bs{4}}\sqb{2\bs{4}}
  + \textrm{permutations}	
  \nn\\
&+f^{abc}\, c_{10}^{(2)}\,\frac{s_{13}-s_{23}}{\Lambda^6}\,
\sqb{13}\sqb{23}\sqb{1\bs{4}}\sqb{2\bs{4}}
 + \textrm{permutations}
\nn\\
  &+	d^{abc}\,c^{(1)}_{12}\,\frac{m^2}{\Lambda^8}\,
	\sqb{13}\sqb{23}\sqb{12}^2\anb{1\bs{4}} \anb{2\bs{4}}
 + \textrm{permutations}\nn\\
        &+
 f^{abc}\,\frac{m\, c_{12}^{\parn 3}}{\Lambda^{8}}\,
 \Big[
 (s_{12}-s_{13})\anb{24}\sqb{24} + (s_{12}-s_{23})\anb{14}\sqb{14}
 \Big]
\sqb{12}\sqb{23}\sqb{31}\,,
 \nn
\end{align}
where we absorbed some signs and a factor of 2 in the
coefficients\footnote{It is easy to see, using~\eqref{massive_spin_1_ppp_n},
  that $c_{12}^{(1)}$ and $c_{12}^{(3)}$ can be absorbed
  in the remaining terms up to pieces of order $m$,
after symmetrizing and using the Schouten identities.}.
It is now straightforward to see the massless limit of this expression.
All we need to do is to absorb $m$ in the scale, and unbold 4-spinors.
The first two lines, with $[14][24]$, give
$\mathcal{M}\parn{Z^{\prime+}; +++}$,
and the third line, with $\anb{14}\anb{24}$,
gives $\mathcal{M}\parn{Z^{\prime-}; +++}$~\footnote{Thus,
  there is no contribution to the
  $\mathcal{M}\parn{Z^{\prime-}; +++}$
  amplitude with $f^{abc}$ up to $d \le 12$.}.
The counting and correspondence to the EFT operators is summarized 
in~\tableref{EFT_X_3g_operator_list}, and reproduces the
results of~\cite{Henning:2015alf}.

Finally, 
the fourth line of~\eqref{zppp_zerom1} 
gives  $\mathcal{M}\parn{Z^{\prime0}; +++}$.
Note that this the lowest order contribution to this amplitude
in $1/\Lambda$.
In particular, the $c_{10}^{(2)}$ term in~\eqref{zppp_zerom1} does not contribute. 
We can rewrite this term as,
\beqa
 f^{abc}\,\frac{m\, c_{12}^{\parn 3}}{\Lambda^{8}}\,
 \Big( s_{12}s_{13}&+&s_{23}s_{13}+s_{12}s_{23}+\mathcal{O}\parn{m^{2}}\Big)
 \sqb{12}\sqb{23}\sqb{31}=\\
 f^{abc}\,\frac{c_{12}^{\parn 3}}{\tilde\Lambda^7}\,
 \Big( s_{12}s_{13}&+&s_{23}s_{13}+s_{12}s_{23}+\Big)
 \sqb{12}\sqb{23}\sqb{31} + \mathcal{O}\parn{m^{2}}
 \,,\nn
\eeqa
which is the same as the dimension-11 contribution
in~\eqref{higgs_3g_all_plus_final_result}.
Unlike in the $Z^\prime; --+$ case, here the longitudinal $Z^\prime$ polarization
does not reproduce the leading contribution in $h;+++$, which appears
at dimension-7, but only the next non-zero contribution,
which is dimension-11.
This is consistent with what we expect based on the Higgs mechanism.
In an EFT operator generating the  $Z^{\prime0};+++$ amplitude we can replace
$Z^\prime_\mu$ by $\partial_\mu h$. Thus the corresponding scalar operator should
contain at least one derivative. The first such operator is $D^4 h G^3$
(see~\tableref{EFT_h_3g_operator_list}), with the color structure $f^{abc}$. 
This is indeed a dimension-11 operator.

To summarize, we have seen how the massive $Z^\prime$ EFT amplitudes
decompose into 
three distinct amplitudes corresponding to the
different $Z^\prime$ polarizations.
In the massless limit, these simply reduce to the
$Z^{\prime+}$-helicity amplitude, the $Z^{\prime-}$-helicity amplitude, and
the scalar amplitude, which supplies the longitudinal $Z^\prime$ polarization.

\section{Conclusions}\label{sec:concl}
We have used on-shell methods to derive tree-level helicity amplitudes involving a
new SM-singlet, of spin zero or one, with couplings to gluons.
Specifically, we calculated $h;ggg$ amplitudes 
and $hh;gg$ amplitudes for a scalar field $h$, which is either massive or massless, up to dimension-13.
We also calculated $Z^\prime;ggg$ amplitudes for a massive vector $Z^\prime$,
up to dimension-12, and showed how these amplitudes decompose into the massless
vector plus scalar amplitudes.

It is straightforward to replace the gluons by photons in our results.
Furthermore, the amplitudes we calculated can be used to obtain  amplitudes with the
gluons replaced by weak gauge bosons in the high-energy limit.
Throughout, we ignored quarks, and it would be interesting to extend this approach to
include fermions, and in particular, to reconstruct the full SM EFT.
The calculation of EFT amplitudes at the loop level,
where the effects of renormalization and operator mixing would show up,
is another obvious direction that we plan to explore.

Our results can be used to calculate the  LHC production and decay of new massive scalars
or vectors coupled to the SM gluons through higher-dimension operators.
In particular, since spin information is preserved, they allow for various spin and
coupling measurements of the new particles.

As we have shown, it is easy to infer the structure of EFT amplitudes,
and thereby count the number of operators of very high dimensionality.
For practical purposes however, the contributions of interest
are the lowest order ones. When interpreting LHC measurements, the EFT results
are reliable only for energies sufficiently smaller than $\Lambda$.
Thus, deriving the structure of amplitudes at dimension-13 is probably
just an academic, if pleasing, exercise.
Still, what on-shell methods clearly reveal is the power of Lorentz symmetry
when applied to physical quantities. 
Gauge-redundancies and off-shell variables are stripped away,
exposing the simplicity of the possible underlying structures of the theory.
In the case of EFTs, operator redundancies are stripped away too.
One may therefore wonder whether an on-shell approach to effective field theories
would ultimately lead us towards a more fundamental theory beyond the standard model.

\section*{Acknowledgments}
We thank Teppei Kitahara for pointing out a sign error in a previous version of this paper, and for useful discussions.
YS thanks the Aspen Center for Physics, supported in part by NSF-PHY-1607611, where
this work was completed.
Research supported by the Israel Science Foundation (Grant No.~720/15),
by the United-States-Israel Binational Science Foundation (BSF) (Grant No.~2014397),
and by the ICORE Program of the Israel Planning and Budgeting Committee (Grant No.~1937/12).


\appendix
\section{Notation}
\label{sec:notation}
We follow the conventions and notation of~\cite{Schwartz:Book2014} for spinors, with the metric 
$\parn{+,-,-,-}$.
The Levi-Civita symbol is given by 
\begin{align}
	&\epsilon^{\alpha\beta} = - \epsilon_{\alpha\beta}= \epsilon^{\dot{\alpha}\dot{\beta}} = - \epsilon_{\dot{\alpha}\dot{\beta}} = 
	\parn{\begin{array}{cc}
		0 & 1\\
		-1 & 0
	\end{array}}\,,
\end{align}
and can be used to raise and lower spinor indices, such that
\begin{align}\label{eqn:raise_lower}
	&\psi^{\alpha} = \epsilon^{\alpha\beta}\psi_{\beta}\,, \quad \psi_{\alpha} = \epsilon_{\alpha\beta}\psi^{\beta}\,, \quad \epsilon^{\alpha\beta}\epsilon_{\beta\gamma} = \delta^{\alpha}_{\ \gamma}
\end{align}
with the same expression for the dotted indices. 

Inside spinor brackets, indices are contracted such that
\begin{align}
	&\anb{\lambda\chi} \equiv \lambda_{\alpha}\chi^{\alpha} = -\lambda^{\alpha}\chi_{\alpha}  \,, \quad \sqb{\lambda\chi} \equiv \tilde{\lambda}^{\dot{\alpha}}\tilde{\chi}_{\dot{\alpha}} =  -\tilde{\lambda}_{\dot{\alpha}}\tilde{\chi}^{\dot{\alpha}} \,.
\end{align}
To simplify the notation we introduce the ``half-bracket'' 
\begin{equation}\label{eqn:spinor_helicity_variables_half_bracket_notation}
  \lambda^{\alpha}=\ket p\,,\quad\lambda_{\alpha}=\bra p\,,\quad\tilde{\lambda}_{\dot{\alpha}}=
  \Ket p\,,\quad\tilde{\lambda}^{\dot{\alpha}}=\Bra p\,,
\end{equation}
such that lightlike momenta are given by 
\begin{equation}\label{eqn:spinor_helicity_variables_momentum_in_half_brackets}
p^{\alpha\dot{\alpha}}=\ket p\Bra p\,,\quad p_{\dot{\alpha}\alpha}=\Ket p\bra p\,,
\end{equation}
where 
\begin{align}
	&p^{\alpha\dot{\alpha}} \equiv \sigma^{\alpha\dot{\alpha}}_{\mu} p^{\mu} \eqCS 
	p_{\dot{\alpha}\alpha} \equiv \overline{\sigma}_{\dot{\alpha}\alpha}^{\mu} p_{\mu} \,,
\end{align}
and $\sigma^{\mu \alpha \dot{\alpha}} = \parn{\delta^{\alpha \dot{\alpha}}, \vec{\sigma}^{\alpha \dot{\alpha}}}$, 
$\bar{\sigma}^{\mu}_{\dot{\alpha} \alpha} = \parn{\delta_{\dot{\alpha} \alpha}, -\vec{\sigma}_{\dot{\alpha} \alpha}}$. 

It is sometimes useful to have explicit expressions for the spinors.
Writing,
$$p^{\mu} = E\parn{1,\cos\phi \sin\theta,\sin\phi \sin\theta,\cos\theta},$$
we have,
\begin{align}\label{eqn:spinor_helicity_variables_in_term_of_momentum}
  \ket{p} = \sqrt{2E} \, \parn{\begin{array}{c} s \\ -c \\
  \end{array}} \eqCS \Bra{p} = \sqrt{2E} \, \parn{s, -c^{*}}\,,
\end{align}
where $s\equiv \sin\parn{\frac{\theta}{2}}$ and $c\equiv \cos\parn{\frac{\theta}{2}}e^{i\phi}$.

Contracting lightlike momenta amounts to tracing over the half-brackets with a $\frac{1}{2}$ factor 
\begin{equation}\label{eqn:spinor_helicity_variables_momentum_contraction}
p\cdot q=p^{\mu}q_{\mu}=\frac{1}{2}p^{\alpha\dot{\alpha}}q_{\dot{\alpha}\alpha}\equiv\frac{1}{2}\tr\left\{ \ket p\left[pq\right]\bra q\right\} =\frac{1}{2}\left\langle qp\right\rangle \left[pq\right]\,.
\end{equation}

The decomposition in \eqref{spinor_helicity_variables_momentum_in_half_brackets} is invariant under 
\begin{align}\label{eqn:spinor_helicity_variables_LG_transf_in_half_brackets}
	&\ket{p} \to \xi \ket{p} \,, \qquad \Bra{p} \to \frac{1}{\xi} \Bra{p}\,,
\end{align}
where for real momenta $\xi$ is just a pure phase. 
\eqref{spinor_helicity_variables_LG_transf_in_half_brackets} is thus a U(1) little group transformation of the lightlike momentum.

The polarizations for a massless vector are given by 
\begin{align}\label{eqn:polarizations_sh_form_pols_vec_spinors}
\parnS{\epsilon^{-}_{p}\parn{r}}^{\alpha\dot{\alpha}} = \sqrt{2}\frac{\ket{p}\Bra{r}}{\parnS{pr}}\,, \qquad \parnS{\epsilon^{+}_{p}\parn{r}}^{\alpha\dot{\alpha}} = \sqrt{2}\frac{\ket{r}\Bra{p}}{\left\langle rp\right\rangle }\,,
\end{align}
where $r$ is a reference momentum not aligned with $p$, \ie, $p \cdot r \ne 0$. 

The polarizations~\eqref{polarizations_sh_form_pols_vec_spinors} are not invariant under the little group
transformation~\eqref{spinor_helicity_variables_LG_transf_in_half_brackets}. 
Thus, a positive helicity gluon carries helicity weight  $-2$, while a negative helicity gluon has helicity weight $+2$. 

Massive momenta can be decomposed in much the same way. 
However, because the determinant is not zero the decomposition will require a pair of massless spinors
\begin{equation}
p^{\alpha\dot{\alpha}}=\psi^{\alpha}\tilde{\psi}^{\dot{\alpha}}+\eta^{\alpha}\tilde{\eta}^{\dot{\alpha}}\,.
\end{equation}
Following~\cite{Arkani-Hamed:2017jhn} we pack the pair into a doublet, such that 
\begin{equation}
\lambda_{I}\equiv\left(\begin{array}{cc}\psi & \eta\end{array}\right) \,, \qquad \tilde{\lambda}^{I}\equiv\left(\begin{array}{c}
\tilde{\psi}\\
\tilde{\eta}
\end{array}\right)\,,
\end{equation}
which enables us to write the massive momentum 
\begin{equation}\label{eqn:spinor_helicity_variables_massive_momentum_in_half_brackets}
p^{\alpha\dot{\alpha}} = \bs{\lambda}^{\alpha}_{I}\tilde{\bs{\lambda}}^{\dot{\alpha}I} \equiv \ket{\bs{p}_{I}}\Bra{\bs{p}^{I}} \,, \quad 
p_{\dot{\alpha}\alpha} = - \tilde{\bs{\lambda}}_{\dot{\alpha} I} \bs{\lambda}^{I}_{\alpha} \equiv -\Ket{\bs{p}_{I}}\bra{\bs{p}^{I}} \,,
\end{equation}
where $I=1,2$ is an $\SUNlr{2}{L}$. 
Note, that for real momenta we have the reality conditions 
\begin{align}
	\bra{\bs{p}^{I}}^{\dagger}=-\Ket{\bs{p}_{I}}\eqCS\Ket{\bs{p}_{I}}^{\dagger}=-\bra{\bs{p}^{I}}\eqCS\ket{\bs{p}_{I}}^{\dagger}=\Bra{\bs{p}^{I}}\eqCS\Bra{\bs{p}^{I}}^{\dagger}=\ket{\bs{p}_{I}}\,.
\end{align} 
The little group transformation now takes the form $\tilde{\lambda}^{I}\rightarrow W_{\ J}^{I}\tilde{\lambda}^{J}\,, \; \lambda_{I}\rightarrow\left(W^{-1}\right)_{\ I}^{J}\lambda_{J}$
where the $W$'s are $\SUN 2$ matrices. 
Squaring a massive momentum gives
\begin{align}
	p^{2} =\frac{1}{2} p^{\alpha\dot{\alpha}} p_{\dot{\alpha}\alpha} = \frac{1}{2} \sqb{\bs{p}^{J} \bs{p}^{I}} \anb{\bs{p}_{I} \bs{p}_{J}} = m^{2}\,.
\end{align}
Expanding the brackets in $\SUNlr{2}{L}$ invariants we get 
\begin{align}\label{eqn:spinor_helicity_variables_massive_momentum_rels_between_mass_and_brackets}
	\sqb{\bs{p}^{J}\bs{p}^{I}}=\tilde{\mathfrak{M}}\epsilon^{IJ}\eqCS 
	\anb{\bs{p}_{I}\bs{p}_{J}}=\mathfrak{M}\epsilon_{JI}\,,
\end{align}
where $\tilde{\mathfrak{M}} \times \mathfrak{M} = m^{2}$ and the $\SUNlr{2}{L}$ indices can be raised and lowered with the Levi-Civita symbol, just as in~\eqref{raise_lower}, with 
$\epsilon^{IJ} = -\epsilon_{IJ}$
and $\epsilon^{12}=1$.
Forbidding opposite rephasing of the mass parameters $\tilde{\mathfrak{M}}$ and $\mathfrak{M}$, which are not little group transformations, we can set $\tilde{\mathfrak{M}} = \mathfrak{M} = m$ as in~\cite{Arkani-Hamed:2017jhn}. 
With the decomposition of the massive momentum~\eqref{spinor_helicity_variables_massive_momentum_in_half_brackets},
we can trade any dotted spinor for an undotted  spinor, and vice versa,
\begin{align}\label{eqn:spinor_helicity_variables_massive_momentum_rels_between_half_brackets}
 	&p\ket{\bs{p}^{J}} = m \Ket{\bs{p}^{J}}\,, \quad   p\Ket{\bs{p}^{J}}= m \ket{\bs{p}^{J}}  \,, \quad   \Bra{\bs{p}_{J}}p = -m\bra{\bs{p}_{J}}     \,, \quad    \bra{\bs{p}_{J}}p =  -m\Bra{\bs{p}_{J}}   \,.                    
\end{align}
For amplitudes in which particles 1, 2 and 3 are massless, while 4 is massive, 
momentum conservation then implies,
\begin{align}\label{eqn:massive_spinor_rels}
	&\anb{12}\sqb{2\bs{4}} + \anb{13}\sqb{3\bs{4}} + m\anb{1\bs{4}} = 0\,,~~~
	\sqb{12}\anb{2\bs{4}} + \sqb{13}\anb{3\bs{4}} + m\sqb{1\bs{4}} = 0\,,
\end{align}
and similarly for permutations of 1, 2, 3.

\section{High energy limit}\label{sec:helimit}
The massive spinor notation relies on splitting the massive momentum into
two lightlike vectors.
To avoid index clutter, let us write $p_4=k+q$, with $k^2=q^2=0$ and
$2k\cdot q= m^2$. Then we can identify $\Ket{4^1}=\Ket{k}$,
$\Ket{4^2}=\Ket{q}$,
$\ket{4^1}=\ket{q}$, $\ket{4^2}=-\ket{k}$.
In the massless limit, without loss of generality, $q\to0$ and $k\to p_4$.
Thus $\ket{4^1},\Ket{4^2}={\mathcal O}(m)$, while
$\ket{4^2},\Ket{4^1}$ are finite.
Consider for example the square bracket $\sqb{14^2}=\sqb{1q}$.
We can rewrite it as follows,
\beq\label{eqn:convert}
\sqb{1q}=\frac1{m}\,[1\vert {p_4} \vert k\rangle =
    \frac1{m}\,\Big(
    \sqb{12}\anb{2k} + \sqb{13}\anb{3k}
\Big)\,,
\eeq
     where in the last step we used momentum conservation.
     This is nothing but~\eqref{massive_spinor_rels}, written in terms of
     the momenta $k$ and $q$.
     Note that there is nothing singular in the massless limit of this
     equation, since the numerator of the RHS is   ${\mathcal O}(m^2)$.
     Thus both sides of~\eqref{convert} are order $m$.

     Similarly,
     \beq
 \sqb{1k}=\frac1{m}\,[1\vert {p_4} \vert q\rangle =
    -\frac1{m}\,\Big(
    \sqb{12}\anb{2q} + \sqb{13}\anb{3q}
\Big)\,.
     \eeq    
Both sides of this equation are finite.

Finally, it is useful to write explicit examples of the various spinors.
Choosing the $z$ axis as the direction of the massive momentum,
$p^{\mu} = \parn{E,0,0,p}$,
we can take,
\begin{align}
	k^{\mu} = \frac{E+p}{2}\parn{1, 0, 0, 1}\eqCS q^{\mu} = \frac{E-p}{2}\parn{1, 0, 0, -1}\,,
\end{align} 
with $p = k + q$. 
In the high energy limit, $E-p\approx \frac{m^{2}}{2E}$.
With the timelike momentum $p$ give by~\eqref{spinor_helicity_variables_massive_momentum_in_half_brackets} we can recover the boldface spinors with their little group
indices 
\begin{align}
	\ket{\bs{p}_{1}} = \sqrt{E+p} \, \parn{\begin{array}{c} 0 \\ 1 \\	\end{array}} \eqCS 
	\ket{\bs{p}_{2}} = \sqrt{E-p} \, \parn{\begin{array}{c} 1 \\ 0 \\	\end{array}} \,,
\end{align} 
where we used~\eqref{spinor_helicity_variables_in_term_of_momentum} and chose the kets such that~\eqref{spinor_helicity_variables_massive_momentum_rels_between_mass_and_brackets} holds. 
Then in the high energy limit, 
\begin{align}
  \ket{\bs{p}^{1}} \eqCS
  \ket{\bs{p}_{2}} \eqCS
  \Bra{\bs{p}^{2}} \eqCS
  \Bra{\bs{p}_{1}} & \sim \mathcal{O} \parn{\frac{m}{\sqrt{E}}}\,,
\end{align}
while
\begin{align}
  \ket{\bs{p}_{1}} \eqCS
  \ket{\bs{p}^2} \eqCS
  \Bra{\bs{p}^{1}} \eqCS
  \Bra{\bs{p}_{2}} & \sim \mathcal{O} (\sqrt{E})\,.
\end{align}
%

\bibliography{Amplitude_EFT}

\providecommand{\href}[2]{#2}\begingroup\raggedright\begin{thebibliography}{10}

\bibitem{Jenkins:2009dy}
E.~E.~Jenkins and A.~V.~Manohar, {\em {Algebraic Structure of Lepton and Quark
  Flavor Invariants and CP Violation}},
  \href{https://dx.doi.org/10.1088/1126-6708/2009/10/094}{JHEP {\bfseries 10}
  (2009) 094}
{\ttfamily [\href{https://arxiv.org/abs/0907.4763}{arXiv:0907.4763}]}.

\bibitem{Grzadkowski:2010es}
B.~Grzadkowski, M.~Iskrzynski, M.~Misiak, and J.~Rosiek, {\em {Dimension-Six
  Terms in the Standard Model Lagrangian}},
  \href{https://dx.doi.org/10.1007/JHEP10(2010)085}{JHEP {\bfseries 10} (2010)
  085}
{\ttfamily [\href{https://arxiv.org/abs/1008.4884}{arXiv:1008.4884}]}.

\bibitem{Lehman:2015coa}
L.~Lehman and A.~Martin, {\em {Low-derivative operators of the Standard Model
  effective field theory via Hilbert series methods}},
  \href{https://dx.doi.org/10.1007/JHEP02(2016)081}{JHEP {\bfseries 02} (2016)
  081}
{\ttfamily [\href{https://arxiv.org/abs/1510.00372}{arXiv:1510.00372}]}.

\bibitem{Henning:2015alf}
B.~Henning, X.~Lu, T.~Melia, and H.~Murayama, {\em {2, 84, 30, 993, 560, 15456,
  11962, 261485, ...: Higher dimension operators in the SM EFT}},
  \href{https://dx.doi.org/10.1007/JHEP08(2017)016}{JHEP {\bfseries 08} (2017)
  016}
{\ttfamily [\href{https://arxiv.org/abs/1512.03433}{arXiv:1512.03433}]}.

\bibitem{Lehman:2015via}
L.~Lehman and A.~Martin, {\em {Hilbert Series for Constructing Lagrangians:
  expanding the phenomenologist's toolbox}},
  \href{https://dx.doi.org/10.1103/PhysRevD.91.105014}{Phys.\  Rev.\
  {\bfseries D91} (2015) 105014}
{\ttfamily [\href{https://arxiv.org/abs/1503.07537}{arXiv:1503.07537}]}.

\bibitem{Henning:2017fpj}
B.~Henning, X.~Lu, T.~Melia, and H.~Murayama, {\em {Operator bases,
  $S$-matrices, and their partition functions}},
  \href{https://dx.doi.org/10.1007/JHEP10(2017)199}{JHEP {\bfseries 10} (2017)
  199}
{\ttfamily [\href{https://arxiv.org/abs/1706.08520}{arXiv:1706.08520}]}.

\bibitem{Gripaios:2018zrz}
B.~Gripaios and D.~Sutherland, {\em {DEFT: A program for operators in EFT}},
{\ttfamily \href{https://arxiv.org/abs/1807.07546}{arXiv:1807.07546}} (2018).

\bibitem{Cheung:2016drk}
C.~Cheung, K.~Kampf, J.~Novotny, C.-H.~Shen, and J.~Trnka, {\em {A Periodic
  Table of Effective Field Theories}},
  \href{https://dx.doi.org/10.1007/JHEP02(2017)020}{JHEP {\bfseries 02} (2017)
  020}
{\ttfamily [\href{https://arxiv.org/abs/1611.03137}{arXiv:1611.03137}]}.

\bibitem{DelDuca:2004wt}
V.~Del~Duca, A.~Frizzo, and F.~Maltoni, {\em {Higgs boson production in
  association with three jets}},
  \href{https://dx.doi.org/10.1088/1126-6708/2004/05/064}{JHEP {\bfseries 05}
  (2004) 064}
{\ttfamily [\href{https://arxiv.org/abs/hep-ph/0404013}{hep-ph/0404013}]}.

\bibitem{Dixon:2004za}
L.~J.~Dixon, E.~W.~N.~Glover, and V.~V.~Khoze, {\em {MHV rules for Higgs plus
  multi-gluon amplitudes}},
  \href{https://dx.doi.org/10.1088/1126-6708/2004/12/015}{JHEP {\bfseries 12}
  (2004) 015}
{\ttfamily [\href{https://arxiv.org/abs/hep-th/0411092}{hep-th/0411092}]}.

\bibitem{Dixon:2013uaa}
L.~J.~Dixon in {\em {Proceedings, 2012 European School of High-Energy Physics
  (ESHEP 2012): La Pommeraye, Anjou, France, June 06-19, 2012}}, pp.~31--67.
\newblock 2014.
\newblock
{\ttfamily \href{https://arxiv.org/abs/1310.5353}{arXiv:1310.5353}}.
\newblock

\bibitem{Dixon:1993xd}
L.~J.~Dixon and Y.~Shadmi, {\em {Testing gluon selfinteractions in three jet
  events at hadron colliders}},
  \href{https://dx.doi.org/10.1016/0550-3213(94)90563-0,
  10.1016/0550-3213(95)00450-7}{Nucl.\  Phys.\  {\bfseries B423} (1994) 3--32}
  {\ttfamily [\href{https://arxiv.org/abs/hep-ph/9312363}{hep-ph/9312363}]}.
[Erratum: Nucl. Phys.B452,724(1995)].

\bibitem{Cheung:2015aba}
C.~Cheung and C.-H.~Shen, {\em {Nonrenormalization Theorems without
  Supersymmetry}},
  \href{https://dx.doi.org/10.1103/PhysRevLett.115.071601}{Phys.\  Rev.\
  Lett.\  {\bfseries 115} (2015) 071601}
{\ttfamily [\href{https://arxiv.org/abs/1505.01844}{arXiv:1505.01844}]}.

\bibitem{Azatov:2016sqh}
A.~Azatov, R.~Contino, C.~S.~Machado, and F.~Riva, {\em {Helicity selection
  rules and noninterference for BSM amplitudes}},
  \href{https://dx.doi.org/10.1103/PhysRevD.95.065014}{Phys.\  Rev.\
  {\bfseries D95} (2017) 065014}
{\ttfamily [\href{https://arxiv.org/abs/1607.05236}{arXiv:1607.05236}]}.

\bibitem{Elvang:Book15}
H.~Elvang and Y.-t.~Huang, {\em {Scattering Amplitudes in Gauge Theory and
  Gravity}}.
\newblock Cambridge University Press,
2015.
\newblock

\bibitem{Cheung:2017pzi}
C.~Cheung, {\em {TASI Lectures on Scattering Amplitudes}},
{\ttfamily \href{https://arxiv.org/abs/1708.03872}{arXiv:1708.03872}} (2017).

\bibitem{Schwartz:Book2014}
M.~D.~Schwartz, {\em Quantum field theory and the standard model.}
\newblock Cambridge : Cambridge University Press, 2014., 2014.

\bibitem{Dawson:2014ora}
S.~Dawson, I.~M.~Lewis, and M.~Zeng, {\em {Effective field theory for Higgs
  boson plus jet production}},
  \href{https://dx.doi.org/10.1103/PhysRevD.90.093007}{Phys.\  Rev.\
  {\bfseries D90} (2014) 093007}
{\ttfamily [\href{https://arxiv.org/abs/1409.6299}{arXiv:1409.6299}]}.

\bibitem{Dawson:2015gka}
S.~Dawson, I.~M.~Lewis, and M.~Zeng, {\em {Usefulness of effective field theory
  for boosted Higgs production}},
  \href{https://dx.doi.org/10.1103/PhysRevD.91.074012}{Phys.\  Rev.\
  {\bfseries D91} (2015) 074012}
{\ttfamily [\href{https://arxiv.org/abs/1501.04103}{arXiv:1501.04103}]}.

\bibitem{Conde:2016vxs}
E.~Conde and A.~Marzolla, {\em {Lorentz Constraints on Massive Three-Point
  Amplitudes}}, \href{https://dx.doi.org/10.1007/JHEP09(2016)041}{JHEP
  {\bfseries 09} (2016) 041}
{\ttfamily [\href{https://arxiv.org/abs/1601.08113}{arXiv:1601.08113}]}.

\bibitem{Arkani-Hamed:2017jhn}
N.~Arkani-Hamed, T.-C.~Huang, and Y.-t.~Huang, {\em {Scattering Amplitudes For
  All Masses and Spins}},
{\ttfamily \href{https://arxiv.org/abs/1709.04891}{arXiv:1709.04891}} (2017).

\bibitem{Witten:2003nn}
E.~Witten, {\em {Perturbative gauge theory as a string theory in twistor
  space}}, \href{https://dx.doi.org/10.1007/s00220-004-1187-3}{Commun.\  Math.\
   Phys.\  {\bfseries 252} (2004) 189--258}
{\ttfamily [\href{https://arxiv.org/abs/hep-th/0312171}{hep-th/0312171}]}.

\bibitem{Simmons:1989zs}
E.~H.~Simmons, {\em {Dimension-six Gluon Operators as Probes of New Physics}},
\href{https://dx.doi.org/10.1016/0370-2693(89)90301-8}{Phys.\  Lett.\
  {\bfseries B226} (1989) 132--136}.

\bibitem{Bramante:2011qc}
J.~Bramante, R.~S.~Hundi, J.~Kumar, A.~Rajaraman, and D.~Yaylali, {\em
  {Collider Searches for Fermiophobic Gauge Bosons}},
  \href{https://dx.doi.org/10.1103/PhysRevD.84.115018}{Phys.\  Rev.\
  {\bfseries D84} (2011) 115018}
{\ttfamily [\href{https://arxiv.org/abs/1106.3819}{arXiv:1106.3819}]}.

\bibitem{Alwall:2012np}
J.~Alwall, M.~Khader, A.~Rajaraman, D.~Whiteson, and M.~Yen, {\em {Searching
  for $Z'$ bosons decaying to gluons}},
  \href{https://dx.doi.org/10.1103/PhysRevD.85.115011}{Phys.\  Rev.\
  {\bfseries D85} (2012) 115011}
{\ttfamily [\href{https://arxiv.org/abs/1202.4014}{arXiv:1202.4014}]}.

\bibitem{Kumar:2012ba}
J.~Kumar, A.~Rajaraman, and D.~Yaylali, {\em {Spin Determination for
  Fermiophobic Bosons}},
  \href{https://dx.doi.org/10.1103/PhysRevD.86.115019}{Phys.\  Rev.\
  {\bfseries D86} (2012) 115019}
{\ttfamily [\href{https://arxiv.org/abs/1209.5432}{arXiv:1209.5432}]}.

\end{thebibliography}\endgroup

\end{document}